\documentclass[fleqn,twoside]{article}
\usepackage{amsmath,amssymb,espcrc2,color,cite}
\usepackage[usenames,dvipsnames]{xcolor}

\usepackage{graphicx,soul}
\usepackage[figuresright]{rotating}

\newcommand{\url}[1]{{\tt #1}}
\newcommand{\lsim}
{\;\raisebox{-.3em}{$\stackrel{\displaystyle <}{\sim}$}\;}
\newcommand{\gsim}
{\;\raisebox{-.3em}{$\stackrel{\displaystyle >}{\sim}$}\;}

\newcommand{\gmt}{\ensuremath{(g-2)_\mu}}

\newcommand{\ssi}{\ensuremath{\sigma^{\rm SI}_p}}

\newcommand{\MZ}{\ensuremath{M_Z}}
\newcommand{\Mh}{\ensuremath{M_h}}

\newcommand{\MA}{\ensuremath{M_A}}

\newcommand{\mgl}{\ensuremath{m_{\tilde g}}}
\newcommand{\msq}{\ensuremath{m_{\tilde q}}}

\newcommand{\sto}[1]{\ensuremath{\tilde t_{#1}}}

\newcommand{\stau}[1]{\ensuremath{\tilde \tau_{#1}}}
\newcommand{\stopone}{\ensuremath{\tilde t_{1}}}

\newcommand{\cha}[1]{\tilde \chi^\pm_{#1}}

\newcommand{\mcha}[1]{\ensuremath{m_{\tilde \chi^\pm_{#1}}}}
\newcommand{\neu}[1]{\tilde \chi^0_{#1}}
\newcommand{\mneu}[1]{\ensuremath{m_{\tilde \chi^0_{#1}}}}

\newcommand{\staue}{\tilde \tau_1}
\newcommand{\mstaue}{m_{\staue}}

\newcommand{\tb}{\ensuremath{\tan\beta}}

\newcommand{\tev}{\ensuremath{\,\, \mathrm{TeV}}}
\newcommand{\gev}{\ensuremath{\,\, \mathrm{GeV}}}
\newcommand{\mev}{\ensuremath{\,\, \mathrm{MeV}}}

\def\reffi#1{\mbox{Fig.~\ref{#1}}}

\def\refse#1{\mbox{Sect.~\ref{#1}}}

\def\citere#1{\mbox{Ref.~\cite{#1}}}

\definecolor{orange}{rgb}{1,0.5,0}

\definecolor{Gray}{named}{Gray}
\newcommand{\gray}[1]{\color{Gray}#1 \color{Black}}

\newcommand{\ETslash}{\ensuremath{/ \hspace{-.7em} E_T}}

\title{\vspace{-1.5cm}
\bf \LARGE Supersymmetric Dark Matter after LHC Run 1 \\ \vspace{0.5em}}

\author{
{\bf E.A.~Bagnaschi}\address[DESY]
   {DESY, Notkestra{\ss}e 85, D--22607 Hamburg, Germany},
{\bf O.~Buchmueller}{\address[Imperial]
   {High\,Energy\,Physics\,Group,\,Blackett\,Laboratory,\,Imperial\,College,\,Prince\,Consort\,Road,\,London\,SW7\,2AZ,\,UK},
\bf R.~Cavanaugh}\address[FNAL]
   {Fermi National Accelerator Laboratory, P.O. Box 500, 
    Batavia, Illinois 60510, USA}\hbox{$^{\rm ,}$}\address[UIC]
   {Physics Department, University of Illinois at Chicago, Chicago, 
    Illinois 60607-7059, USA},
{\bf M.~Citron}\addressmark[Imperial],
{\bf A.~De~Roeck}\address[CERN]
   {Physics Department, CERN, CH--1211 Geneva 23, Switzerland}\hbox{$^{\rm ,}$}\address[Antwerpen]
   {Antwerp University, B--2610 Wilrijk, Belgium},
 {\bf M.J.~Dolan}\address[SLAC]
{Theory Group, SLAC National Accelerator Laboratory,
2575 Sand Hill Road, Menlo Park, \\ CA 94025-7090, USA \&
ARC Centre of Excellence for Particle Physics at the Terascale, School of Physics, University of Melbourne, 3010, Australia},
{\bf J.R.~Ellis}\address[KCL]{Theoretical Particle Physics
  and Cosmology Group, Department of Physics, King's College London, London~WC2R~2LS, UK}\hbox{$^{\rm ,}$}\addressmark[CERN], 
{\bf H.~Fl\"acher}\address[Bristol]
   {H.H.~Wills Physics Laboratory, University of Bristol, Tyndall Avenue, Bristol BS8 1TL, UK},
{\bf S.~Heinemeyer}\address[Santander]
   {Instituto de F\'{\i}sica de Cantabria (CSIC-UC), 
    E--39005 Santander, Spain},
{\bf G.~Isidori}\address[Zurich]
{Physik-Institut, Universit\"at Z\"urich, CH-8057 Z\"urich, Switzerland},
{\bf S.~Malik}\addressmark[Imperial],
{\bf D.~Mart\'inez~Santos}\address[USdC]{Universidade de Santiago de Compostela, 
E-15706 Santiago de Compostela, Spain},
{\bf K.A.~Olive}\address[Minnesota] 
{William I.\ Fine Theoretical Physics Institute, School of Physics and
 Astronomy, University of Minnesota, Minneapolis, Minnesota 55455, USA}, 
{\bf K.~Sakurai}\addressmark[KCL],
{\bf K.J.~de~Vries}\addressmark[Imperial],
{\bf G.~Weiglein}\addressmark[DESY]
}

\begin{document}
\begin{abstract}
\vspace{0.5cm}

Different mechanisms operate in various regions of the MSSM parameter
space to bring the relic density of the lightest neutralino, $\neu1$, assumed here
to be the LSP and thus the Dark Matter (DM) particle, into the range
allowed by astrophysics and cosmology. 
These mechanisms include coannihilation with some nearly-degenerate next-to-lightest supersymmetric particle (NLSP)
such as the lighter stau $\stau1$, stop $\stopone$ or chargino $\cha1$, resonant annihilation via
direct-channel heavy Higgs bosons $H/A$, {the light Higgs boson $h$ or the $Z$ boson,} and enhanced annihilation via a larger Higgsino
component of the LSP in the focus-point region. These mechanisms typically select lower-dimensional
subspaces in MSSM scenarios such as the CMSSM, NUHM1, NUHM2 and pMSSM10.
We analyze how future LHC and direct DM searches can complement 
each other in the exploration of the different DM mechanisms within these scenarios.
We find that the ${\staue}$ coannihilation regions of the CMSSM, NUHM1, NUHM2 can largely be explored at the LHC
via searches for $\ETslash$ events and long-lived charged particles, whereas their $H/A$ funnel,
focus-point and $\cha1$ coannihilation regions can largely be explored
by the LZ and Darwin {DM direct detection}
experiments. {We find that} the dominant DM mechanism in our pMSSM10 analysis is
$\cha1$ coannihilation: {parts of its parameter space can be explored by the LHC,
and a larger portion by future direct DM searches}.

\vspace{0.5cm}
\begin{center}
{\tt KCL-PH-TH/2015-33, LCTS/2015-24, CERN-PH-TH/2015-167 \\
DESY 15-132, FTPI-MINN-15/36, UMN-TH-3445/15, SLAC-PUB-16350, FERMILAB-PUB-15-333-CMS}
\end{center}

\end{abstract}

\newpage

\maketitle


\section{Introduction}
\label{sec:intro}

The density of cold dark matter (CDM) in the Universe is now very
tightly constrained, 
in particular by measurements of the cosmic microwave background radiation,
which yield $\Omega_{\rm CDM} h^2 = 0.1186 \pm 0.0020$~\cite{Planck15} and
are consistent with other, less precise, determinations. This determination of
the CDM density at the percent level imposes a corresponding
constraint on the parameters of any model that provides the dominant fraction
of the CDM density. This is, in particular, true for
supersymmetric (SUSY) models
with conserved $R$-parity in which the CDM is provided by the stable
lightest SUSY particle (LSP)~\cite{EHNOS}. In a series of recent papers
incorporating the data from LHC Run~1 and elsewhere, we have
implemented the dark matter (DM) density constraint in global analyses of the parameter
spaces of different variants of the minimal SUSY extension of the Standard
Model (MSSM), assuming that the LSP is the lightest neutralino $\neu1$.
The models studied included the constrained MSSM (CMSSM) with universal soft
SUSY-breaking parameters {($m_0, m_{1/2}$ and $A_0$, in standard notation)} at the GUT scale~\cite{MC9}, the NUHM1(2) in which
universality is relaxed for both together (each separately) of the soft SUSY-breaking
contributions to the masses-squared of the Higgs multiplets
{$m_{H_{1,2}}^2$}~\cite{MC9,MC10}, and {a version of} the
pMSSM10~\cite{MC11}, in 
which 10 of the effective Lagrangian parameters {(3 gaugino masses
  $M_{1,2,3}$, 2 squark masses $m_{\tilde q_{1,2}} \ne m_{\tilde q_{3}}$, a common slepton mass $m_{\tilde \ell}$,
a common trilinear coupling $A_0$, the Higgs mixing parameter $\mu$, the pseudoscalar Higgs mass $\MA$, and the ratio of Higgs vevs $\tb$)} are treated as independent
inputs specified at {the electroweak}
scale.

Reproducing correctly the cosmological CDM density
requires, in general, some special choice of the SUSY model parameters, which
may be some particular combination of sparticle masses and/or couplings.
Examples of the former include hypersurfaces in the SUSY parameter space where the LSP is almost
degenerate in mass with some next-to-lightest SUSY particle (NLSP),
such as the lighter stau $\stau1$~\cite{stauco,Citron:2012fg}, stop $\stopone$~\cite{stopco} or chargino $\cha1$~\cite{chaco}, or where
$\mneu1$ is almost half the mass of a boson such as a heavy Higgs $H/A$ \cite{funnel}, {a light Higgs $h$ or $Z$} \cite{hfunnel}, in
which case rapid direct-channel annihilation may bring the CDM density
into the allowed range. Examples of special coupling combinations include the
focus-point region~\cite{fp}, where the LSP acquires a significant Higgsino component.

We have commented in our previous work on the relevances of these DM
mechanisms for our global analyses. Here we discuss systematically which
DM mechanisms are dominant in which subspaces of the CMSSM~\cite{CMSSM}, NUHM1~\cite{NUHM1,eosknuhm},
NUHM2~\cite{NUHM2,eosknuhm} and pMSSM10~\cite{pMSSM10} parameter spaces, what are the corresponding
experimental signatures, and how one might discover SUSY in each of these
different DM regions.

{Our analysis of the possible detectability of supersymmetry in the CMSSM,
NUHM1, NUHM2 and pMSSM10, depending on the dominant DM mechanisms, is
summarized in Table~\ref{tab:detectability}.}

\section{{Measures of Mass Degeneracy}}

We first introduce measures {on the MSSM parameters that quantify the
relevant mass degeneracies
and define each of the above-mentioned subspaces in the CMSSM, NUHM1 and NUHM2~\cite{MC10,KdV}:}
\begin{align}
{\staue} {\rm ~coann.~(pink):} \hspace{06mm} 
\left(\frac{\mstaue}{\mneu{1}} - 1 \right)   & \,<\,  0.15 \, , \nonumber \\[.3em]
\cha{1} {\rm ~coann.~(green):} \hspace{02mm} 
\left( \frac{\mcha{1}}{\mneu{1}} - 1 \right)  & \,<\, 0.1 \, , \nonumber \\[.3em]
{\tilde t_1} {\rm ~coann.~(grey):} \hspace{06mm} 
\left( \frac{m_{\tilde t_1}}{\mneu{1}} \right) - 1  & \,<\, 0.2 \, , \nonumber \\[.3em]
A/H {\rm ~funnel~(blue):} \hspace{05mm} 
\left|\frac{\MA}{\mneu{1}} - 2 \right|  & \,<\,  0.4 \, , \nonumber \\[.3em]
{\rm focus~point~(cyan):} \hspace{03mm} 
\left( \frac{\mu}{\mneu1} \right) - 1 & \,<\, 0.3 \, .  
\label{shadings}
\end{align}
In each case we also indicate the colour coding we use in the subsequent figures.
The measures (\ref{shadings}) that we use are empirical, but we have verified extensively that
CMSSM, NUHM1 and NUHM2 points {that satisfy the DM density
constraint do} fulfill {at least one of} these conditions, and that they indeed correspond 
to the {dominant} DM mechanisms {(in the sense of giving the largest fractions of final states,
generally $\gtrsim 50$\%)}~\cite{MC10,KdV}. 
We have found that there are some `hybrid' regions {where the dominant mechanism requires two of these conditions
simultaneously.}
In particular, there are regions where {the dominant DM mechanism is $\staue^+ \staue^- \to {\bar b} b$
or ${\bar t} t$, processes involving both stau coannihilation and annihilation via the $A/H$ funnel}, which
we colour purple. There are also regions where the chargino coannihilation condition is
satisfied simultaneously with the stau coannihilation or $A/H$ funnel condition. {However, a
dedicated study using {\tt MicrOMEGAs}~\cite{MicroMegas} shows
that chargino coannihilation is the dominant DM mechanism in these
regions, and that hybrid processes dependent on the $\mcha{1}$ and some other degeneracy conditions being valid
simultaneously are unimportant, so we colour these regions {the same green as the other chargino coannihilation regions.}

The above DM mechanism conditions {need to be modified for our analysis of the pMSSM10. First,} as we see later,
funnels due to annihilations via direct-channel $h$ and $Z$ resonances
can be important~\cite{hfunnel}, so {for the pMSSM10} 
we add to (\ref{shadings}) the supplementary criteria:
\begin{align}
h {\rm ~funnel~(magenta):} \hspace{4mm}
\left|\frac{\Mh}{\mneu{1}} - 2 \right| & \,<\,  0.4 \, , \nonumber \\[.3em]
Z {\rm ~funnel~(orange):} \hspace{4mm}
\left|\frac{\MZ}{\mneu{1}} - 2 \right| & \,<\,  0.4 \, . 
\label{shadings2}
\end{align}
{Secondly, we find that chargino coannihilation dominates in the pMSSM10 also when
the second condition in (\ref{shadings}) is relaxed: we use later the condition
$|\mcha{1}/\mneu{1} - 1|  < 0.25$ in our subsequent analysis, which reproduces better the domains of
dominance by $\cha{1}$ coannihilation}~\footnote{{This approach yields results similar to those of~\cite{KdV},
where empirical constraints combining the masses and neutralino mixing matrix elements are used.}}. 
Finally, we recall that the focusing property of the RGEs is not relevant in the pMSSM10. However, the LSP annihilation rate may
still be enhanced when the fifth measure in (\ref{shadings}) is satisfied, due to a larger Higgsino
component in the LSP, though we find that the dominant DM mechanism in the pMSSM10 generally does not involve
this property}. d{We use the same cyan colour to identify regions where this condition
is satisfied, though it is not due to focus-point behaviour.}

Our discussion here of DM mechanisms is based on our previously-published global likelihood 
analyses of the CMSSM and NUHM1~\cite{MC9}, the NUHM2~\cite{MC10} and the pMSSM10~\cite{MC11}.
The reader wishing to know details of our treatments of the various experimental, phenomenological,
theoretical and cosmological constraints, as well as our strategies for scanning the parameter spaces of
these models is referred to~\cite{MC9,MC10,MC11,olderMC}.


\section{Dominant Dark Matter Mechanisms}
\label{sec:DDMM}

In this section we discuss the various mechanisms that play dominant r\^oles in bringing the
relic density into the experimentally measured interval in our four models.
We display in Fig.~\ref{fig:m0m12} $(m_0, m_{1/2})$ planes for the
CMSSM (upper left)~\cite{MC9}, the NUHM1 (upper right)~\cite{MC9} and
the NUHM2 (lower left)~\cite{MC10}{, while for the pMSSM10 we show
the $(\msq, \mneu1)$ plane (lower right)}, where $\msq$ denotes the
mass of the squarks of the first two generations, which we assume to be common~%
\footnote{{In
some cases, these and subsequent figures may include small updates from the versions shown previously~\cite{MC9,MC10,MC11}, as they
incorporate the latest implementations of the experimental
constraints.}}%
. The $\Delta \chi^2 = 2.30$ and 5.99 contours that we found in global fits
to these models, corresponding approximately to the 68 and 95\% CL contours, are shown as
solid red and blue lines, respectively. {Here and elsewhere, the green
  stars indicate the best-fit points}, {whose exact locations in some parameter planes are poorly determined and do not
  carry much useful information, in general, as the $\chi^2$ minima are quite shallow.}
Also shown, as {solid purple lines, is the current 95\% CL {CMSSM} exclusion from $\ETslash$ searches at the LHC~\footnote{{As
discussed in~\cite{MC9,MC10}, this exclusion curve can be applied to the NUHM1 and NUHM2, also
in the range of $m_0 < 0$ shown in these plots, where we interpret negative $m_0 = {\rm Sign}(m_0^2) \sqrt{m_0^2}$.}}.
The dashed purple contours in the CMSSM, NUHM1 and NUHM2 cases show the prospective 5-$\sigma$
discovery reaches for $\ETslash$ searches at the LHC with 3000/fb at 14~TeV, corresponding approximately to the 95\% CL exclusion
sensitivity with 300/fb at 14~TeV{. In} the pMSSM10 case the dashed purple contour shows the 95\% CL exclusion sensitivity
of the LHC with 3000/fb assuming $\mgl \gg \msq$, and the dash-dotted lines
bound the corresponding sensitivity region assuming $\mgl = 4.5 \tev$. All the planes are colour-coded as listed in 
(\ref{shadings} \ref{shadings2}),
{with regions where none of these processes are dominant left uncoloured}.

\begin{figure*}[htb!]
\vspace{0.5cm}
\begin{center}
\resizebox{7.5cm}{!}{\includegraphics{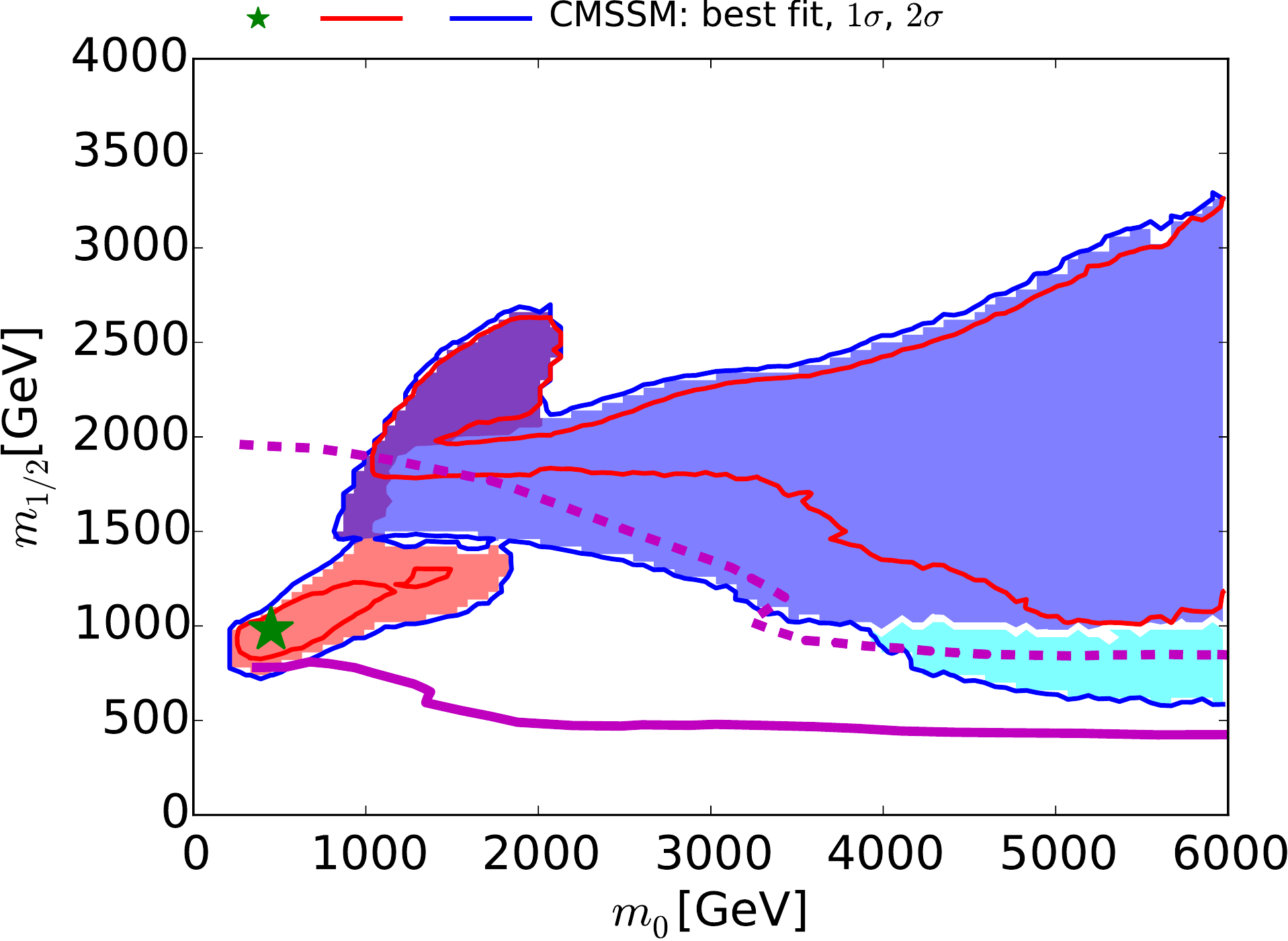}}
\resizebox{7.5cm}{!}{\includegraphics{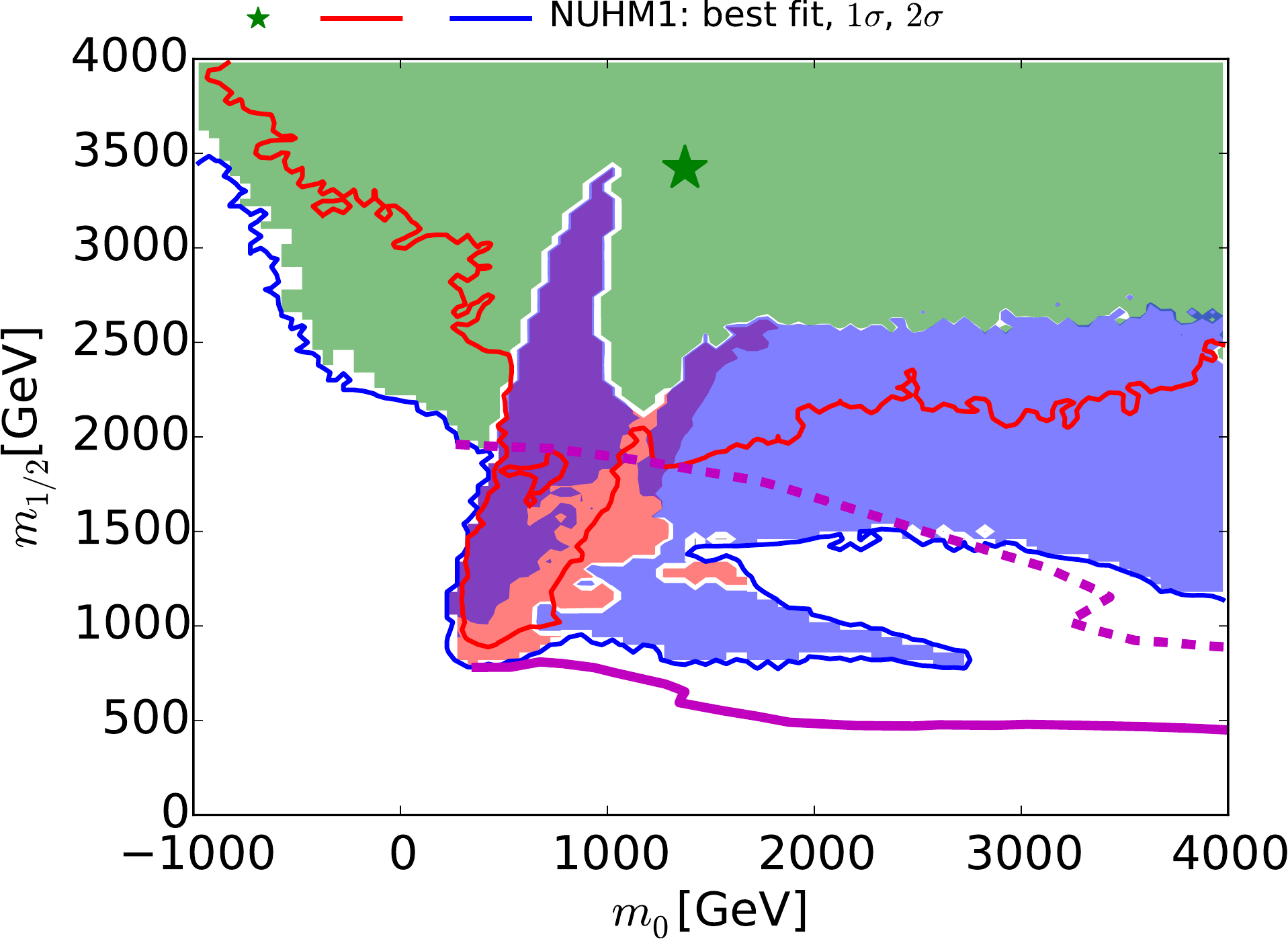}}\\[1em]
\resizebox{7.5cm}{!}{\includegraphics{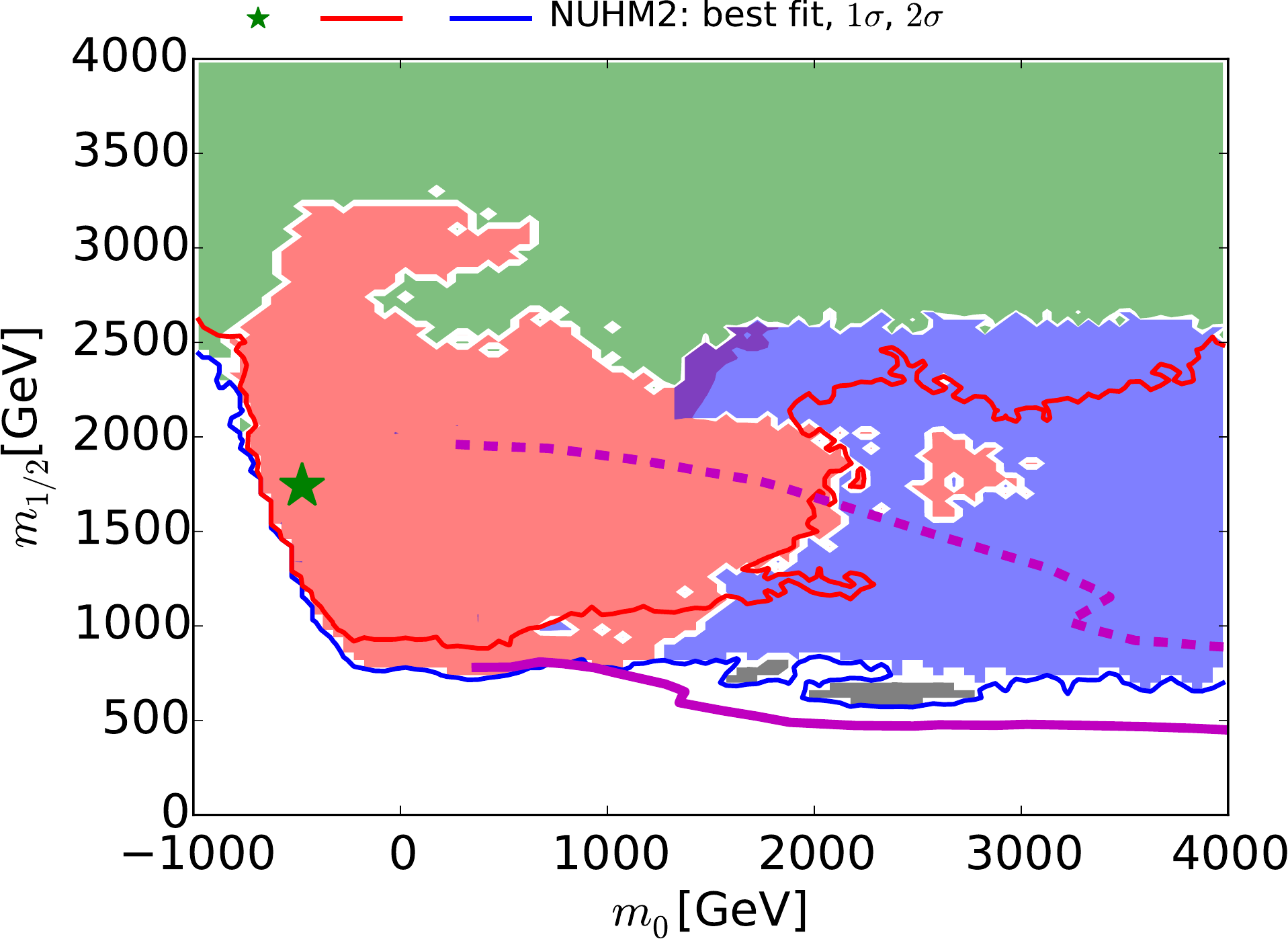}}
\resizebox{7.5cm}{!}{\includegraphics{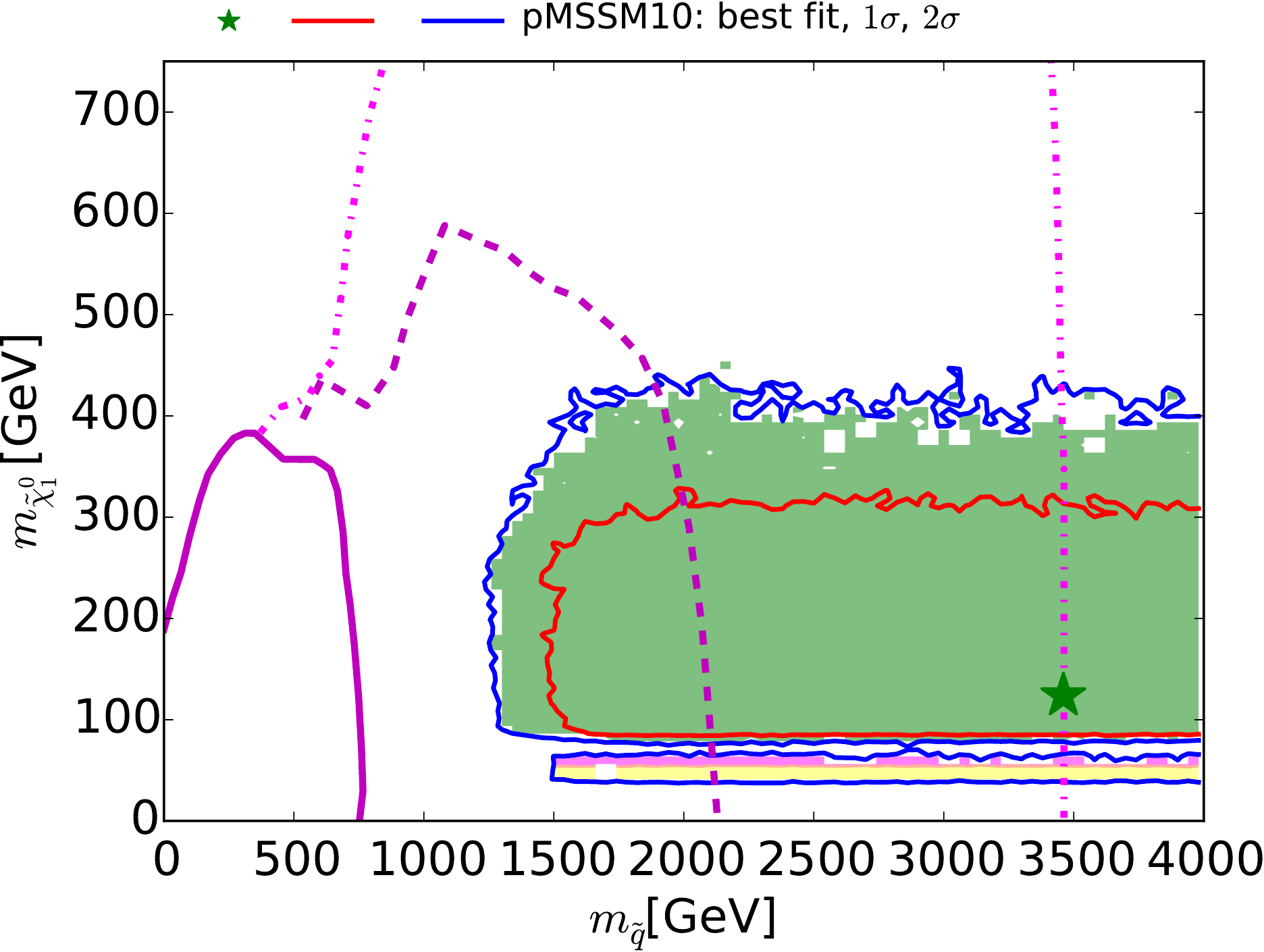}} \\
\resizebox{15cm}{!}{\includegraphics{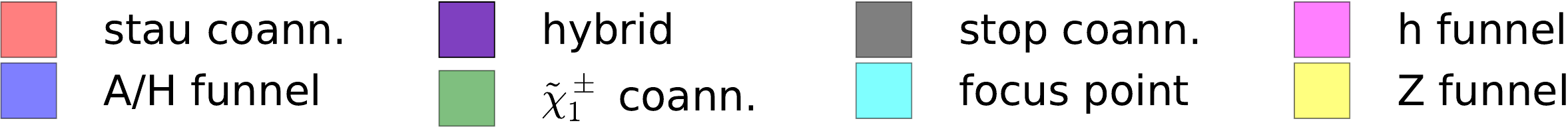}}
\end{center}
\caption{\it The $(m_0, m_{1/2})$ planes in the CMSSM (upper left), the NUHM1 (upper right) and the NUHM2 (lower left),
and the $(\msq, \mneu1)$ plane in the pMSSM10. Regions in which different mechanisms bring the CDM
density into the allowed range are shaded as described in the legend and discussed in the text. The red and blue
contours are the $\Delta \chi^2 = 2.30$ and 5.99 contours found in global fits
to these models, corresponding approximately to the 68 and 95\% CL
contours, with the green stars indicating the best-fit points, and the solid purple contours
show the current LHC 95\% exclusions from $\ETslash$ searches.
{In the CMSSM, NUHM1 and NUHM2 cases, the dashed purple contours show the prospective 5-$\sigma$
discovery reaches for ~$\ETslash$ searches at the LHC with 3000/fb at 14~TeV, corresponding approximately to the 95\% CL exclusion
sensitivity with 300/fb at 14~TeV. In the pMSSM10 case, the dashed purple contour shows the 95\% CL exclusion sensitivity
of the LHC with 3000/fb assuming $\mgl \gg \msq$, and the dash-dotted lines 
bound the corresponding sensitivity region assuming $\mgl = 4.5 \tev$.}
}
\label{fig:m0m12}
\end{figure*}

We see in the upper left panel of Fig.~\ref{fig:m0m12} that three DM mechanisms dominate in the CMSSM:
${\staue}$ coannihilation at low $m_0 \lesssim 2000 \gev$, the $H/A$ funnel at larger $m_0$ and $m_{1/2}$,
and the focus point at larger $m_0$ and smaller $m_{1/2}$. There is also a hybrid ${\staue}/A/H$ region
extending up to $(m_0, m_{1/2}) \sim (2000, 2500) \gev$. In the case of the NUHM1
shown in the upper right panel of Fig.~\ref{fig:m0m12}, {there is an analogous
${\staue}$ coannihilation region. However, it has} a much larger {hybrid ${\staue}/A/H$ region}, 
which has an extension to low $m_{1/2} \sim 1000 \gev$ for $m_0 \lesssim 3000 \gev$.
On the other hand, $\cha{1}$ coannihilation dominates in a large region with $m_{1/2} \gsim 2500 \gev$.
{Here $\mu\ll M_1$, so that the $\neu1, \neu2$ and $\cha1$ are nearly degenerate in mass and
the $\neu1$ has mainly a Higgsino composition.}
A similar $\cha{1}$ coannihilation region is visible in the NUHM2 in the lower left panel of Fig.~\ref{fig:m0m12},
where we also see a more extensive ${\staue}$ coannihilation region
extending to $m_0 \sim 2000 \gev$~\footnote{{Analyses with {\tt micrOMEGAs} confirm that $\cha{1}$ coannihilation
processes contribute over 50\% of the final states in most of the green shaded regions for the NUHM1 and NUHM2, and
generally over 75\% for $m_{1/2} \gtrsim 3000 \gev$.
In the NUHM1 and NUHM2 for $m_{1/2} \lesssim 1500 \gev$, the $\staue$
coannihilation criterion is also satisfied (and that region was colored hybrid in~\cite{MC10}),
but $\staue$ coannihilation does not contribute significantly to the relic density calculation, so here we colour it green.}},
with a relatively {small hybrid ${\staue}/A/H$ region}. This is the only case where we see a region of
dominance by ${\tilde t_1}$ coannihilation, in islands around $(m_0,
m_{1/2}) \sim (2000, 500) \gev$.

In contrast, as shown in
the lower right panel of \reffi{fig:m0m12},
we found in our {version} of the pMSSM10~\cite{MC11} that
the dominant DM mechanism is usually $\cha{1}$ coannihilation, 
{this time with a Bino-like LSP and $M_1\sim M_2$.}~\footnote{{We note,
however, that this is the result of an interplay of various constraints, 
in particular the pMSSM10 assumption of universal slepton masses, as discussed in detail in \citere{MC11}. 
A different selection of pMSSM parameters could favour
regions with different dominant DM mechanisms, in general. However, this is left
for future analysis, and here we simply take over the results
of~\citere{MC11}.}} 
We also note {in \reffi{fig:m0m12}} bands at low $\mneu1$ where rapid
annihilations via the $h$~and $Z$~funnels are dominant. {Not shown in
\reffi{fig:m0m12} are scatterings of points with $\mneu{1} \gtrsim 300 \gev$
where $\staue$ coannihilation can also be important, and of points with
$\mneu{1} \lesssim 150 \gev$ where the fifth condition in (\ref{shadings})
comes into play~\footnote{{Another {\tt micrOMEGAs} analysis shows that $\cha{1}$ coannihilation
processes contribute over 50\% of the final states in most of the green shaded region, and
generally over 75\% for $\mneu{1} \lesssim 250 \gev$.}}. We see later that {this condition and} the $\staue$ coannihilation mechanism
dominate in specific regions of other projections of the pMSSM10 parameter space,
as does `bulk' $\neu{1} \neu{1}$ annihilation where none of the conditions (\ref{shadings}, \ref{shadings2})
are satisfied.}

One may also consider
the possibility that the LSP provides only a fraction of the CDM. A complete discussion of this possibility is
beyond the scope of this paper, but we note that in some regions, e.g., those dominated by $\staue$ or ${\tilde t_1}$
coannihilation, lowering the CDM density requires reducing the NLSP-LSP mass difference. This would also have the effect of
reducing correspondingly the maximal LSP mass, which would favour sparticle detection at the LHC. On the other hand,
direct detection would be more difficult if only a small fraction of the galactic halos were composed of LSPs. For a recent discussion,
see~\cite{Badziak}.


\section{The LHC Sensitivity}

In this section we discuss the prospective reaches of future LHC searches and
their impacts in the contexts of the various preferred DM mechanisms.
{We will see that in many cases the preferred
DM mechanism can be directly probed via the appropriate LHC searches,
as summarized in Table~\ref{tab:detectability}.}

\begin{table*}[htb!]
	\begin{center}
	{
	\begin{tabular}{ | c || c || c | c | c | c|}
		\hline
		DM & Exp't & \multicolumn{4}{c|}{Models} \\ 
		mechanism & & CMSSM & NUHM1 & NUHM2 & pMSSM10 \\ \hline
		${\staue}$ & LHC & {$\checkmark$ $\ETslash$, $\checkmark$ LL} & ($\checkmark$ $\ETslash$, $\checkmark$ LL) & ($\checkmark$ $\ETslash$, $\checkmark$ LL) & ($\checkmark$ $\ETslash$), $\times$ LL \\ 
		coann. & DM & ($\checkmark$) & ($\checkmark$) & $\times$ & $\times$ \\ \hline
		$\cha{1}$ & LHC & -- & $\times$ &  $\times$ & ($\checkmark$ $\ETslash$)  \\ 
		coann. & DM & -- & $\checkmark$ & $\checkmark$ & ($\checkmark$)  \\ \hline
		${\tilde t_1}$ & LHC & -- & -- & $\checkmark$ $\ETslash$ & --  \\ 
		coann. & DM & -- & -- & $\times$ & --  \\ \hline
		$A/H$ & LHC & $\checkmark$ $A/H$ & ($\checkmark$ $A/H$) & ($\checkmark$ $A/H$) & --  \\
		funnel & DM & $\checkmark$ & $\checkmark$ & ($\checkmark$) & -- \\ \hline
		Focus & LHC & ($\checkmark$ $\ETslash$) & -- & -- & --  \\
		point & DM & $\checkmark$ & -- & -- & -- \\ \hline
		$h,Z$ & LHC & -- & -- & -- & ($\checkmark$ $\ETslash$) \\
		funnels & DM & -- & -- & -- & ($\checkmark$) \\ \hline
			\end{tabular}}
			\end{center}
	\caption{\it {
	Summary of the detectability of supersymmetry in the CMSSM, NUHM1, NUHM2 and pMSSM10
	models at the LHC in searches for $\ETslash$ events, long-lived charged particles (LL) and heavy $A/H$
	Higgs bosons, and in direct DM search experiments, depending on the dominant mechanism for
	bringing the DM density into the cosmological range. The symbols $\checkmark$, ($\checkmark$) and
	$\times$ indicate good prospects, interesting possibilities and poorer prospects, respectively.
	The symbol -- indicates that a DM mechanism is not important for the corresponding model.
	{The LHC information is drawn largely from Figs.~\ref{fig:m0m12}, \ref{fig:metastable} and \ref{fig:msqmgl},
	and the direct DM search information from Fig.~\ref{fig:ssi}}.}}
	\label{tab:detectability}
\end{table*}

\subsection{\boldmath{$\ETslash$} Searches}

Looking now at the physics reach of the LHC with $\ETslash$ searches~\cite{LHCsusy}, we see that
in the CMSSM the preferred $\staue$ stau coannihilation region lies just outside the current {LHC 95\% CL 
exclusion region (solid purple line). On the other hand, the $A/H$-funnel  region allowed in the CMSSM at the 95\%
CL lies well outside this region. However, this is no longer the
case in the NUHM1 and particularly the NUHM2, where portions of the $A/H$-funnel  region
lie much closer to the LHC 95\% CL 
exclusion. This is possible} mainly because the Higgs mass constraint is less restrictive
in these models. 
In the CMSSM, the future LHC sensitivity estimated in Fig.~1 of~\cite{Interplay} (dashed purple line)
covers the region where ${\staue}$ coannihilation is dominant, and a part of the hybrid region. It also covers slices of the
$H/A$ funnel region and of the focus-point region. Similar features are seen in the NUHM1, except that
the focus-point region is not visible in this case, but the LHC $\ETslash$ search has no sensitivity in the
$\cha{1}$ coannihilation region. In the case of the NUHM2, the $\ETslash$ search is sensitive to only part
of the ${\staue}$ coannihilation, and none of the $\cha{1}$ coannihilation region, but it does cover
part of the $H/A$ funnel region and all the ${\tilde t_1}$ coannihilation
region. In the pMSSM10 case, parts of the $\cha{1}$ coannihilation region
and the low-$\mneu1$ band are accessible to future LHC searches~%
\footnote{{In principle, these bands are accessible to precise searches for invisible $Z$ and $h$
decays. However, we found in a survey of the pMSSM10 $(\msq, \mneu{1})$ plane in Fig.~\ref{fig:m0m12}
that the $Z \to \neu{1} \neu{1}$ branching ratio exceeds the current experimental uncertainty of $1.5$~MeV
for only a handful of the lowest-$\chi^2$ points. Likewise, the $h \to \neu{1} \neu{1}$ branching ratio exceeds 0.1
also for just a handful of points. Invisible $Z$ and $h$ decays both present opportunities for future searches}. }%
, though the future LHC $\ETslash$ searches 
are less sensitive if $\mgl \gg \msq$ (dashed line) than if 
$\mgl = 4.5 \tev$ (dotted line). {Table~\ref{tab:detectability} summarizes the observability of
$\ETslash$ events that we estimate in the different models considered, depending on the dominant
DM mechanism in each case.}


\subsection{The Possibility of a Long-Lived Charged Sparticle}}

{In some circumstances, a charged sparticle such as the ${\staue}$ or the $\cha{1}$
may be the NLSP, and be only slightly more massive than the LSP, so that it can in principle be long-lived.
As we now show, the presence of a long-lived ${\staue}$ is indeed a distinctive prospective
signature in the CMSSM, NUHM1 and NUHM2.}

\begin{figure*}[htb!]
\begin{center}
\resizebox{7.5cm}{!}{\includegraphics{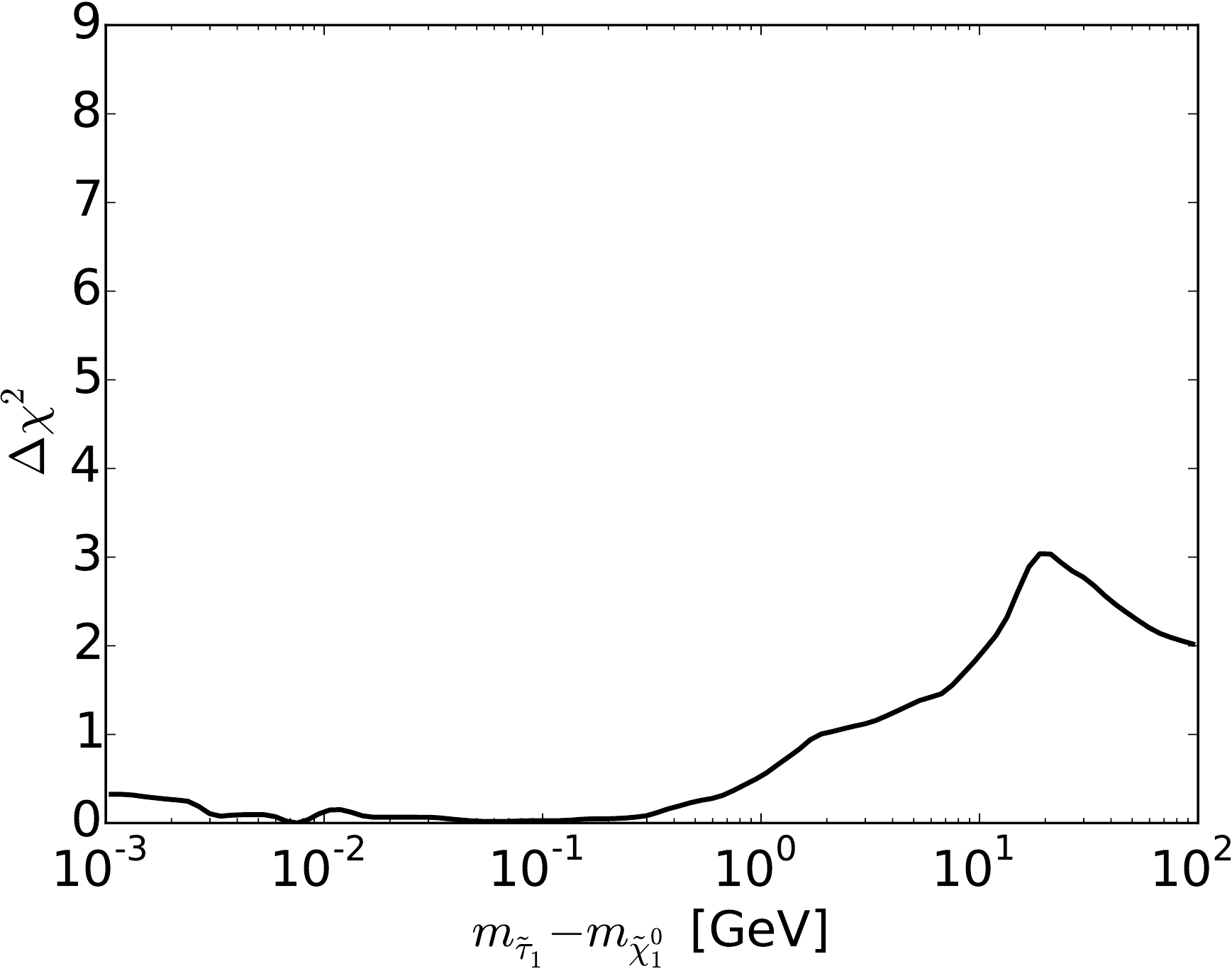}}
\resizebox{7.5cm}{!}{\includegraphics{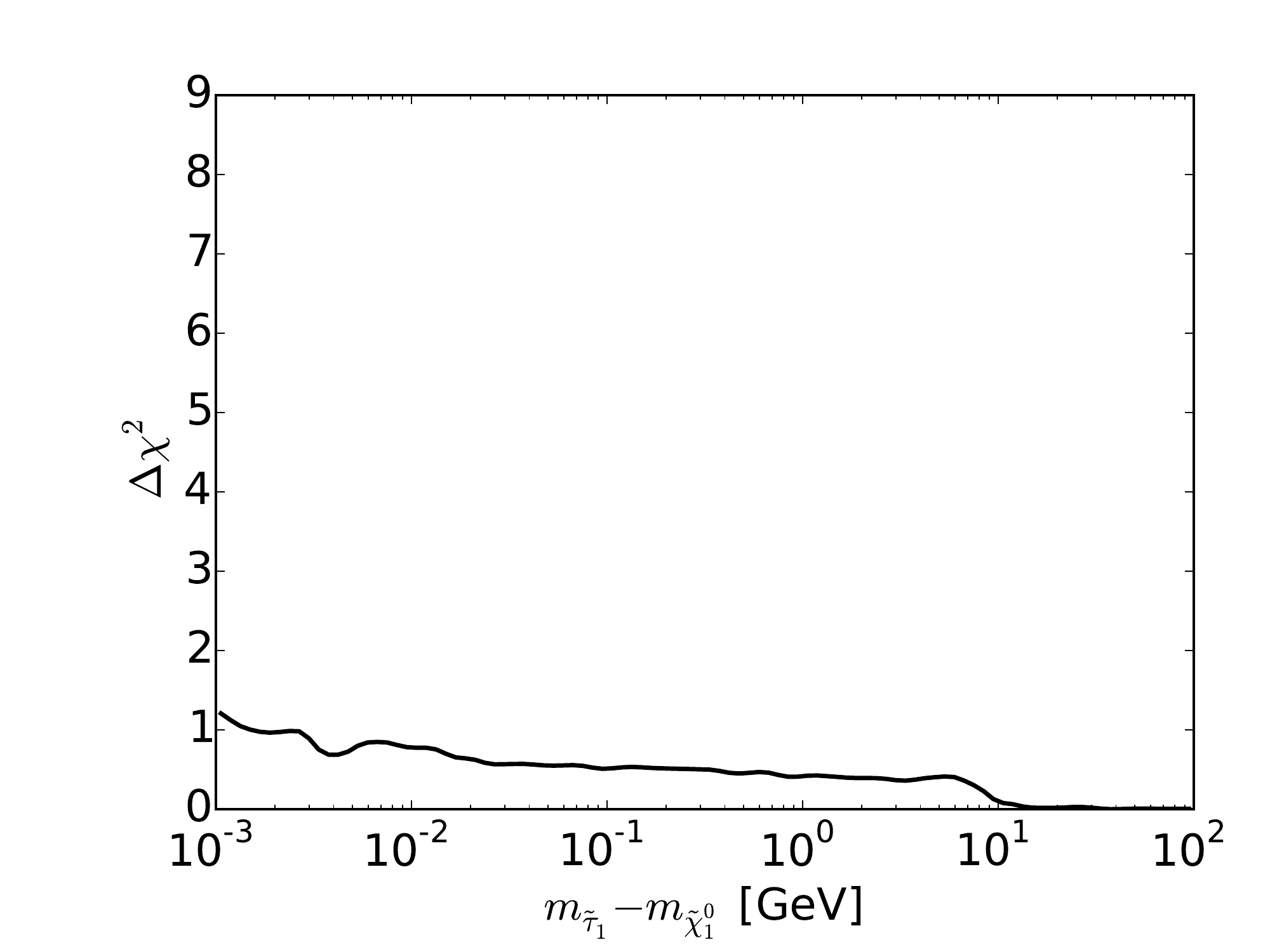}}\\[1em]
\resizebox{7.5cm}{!}{\includegraphics{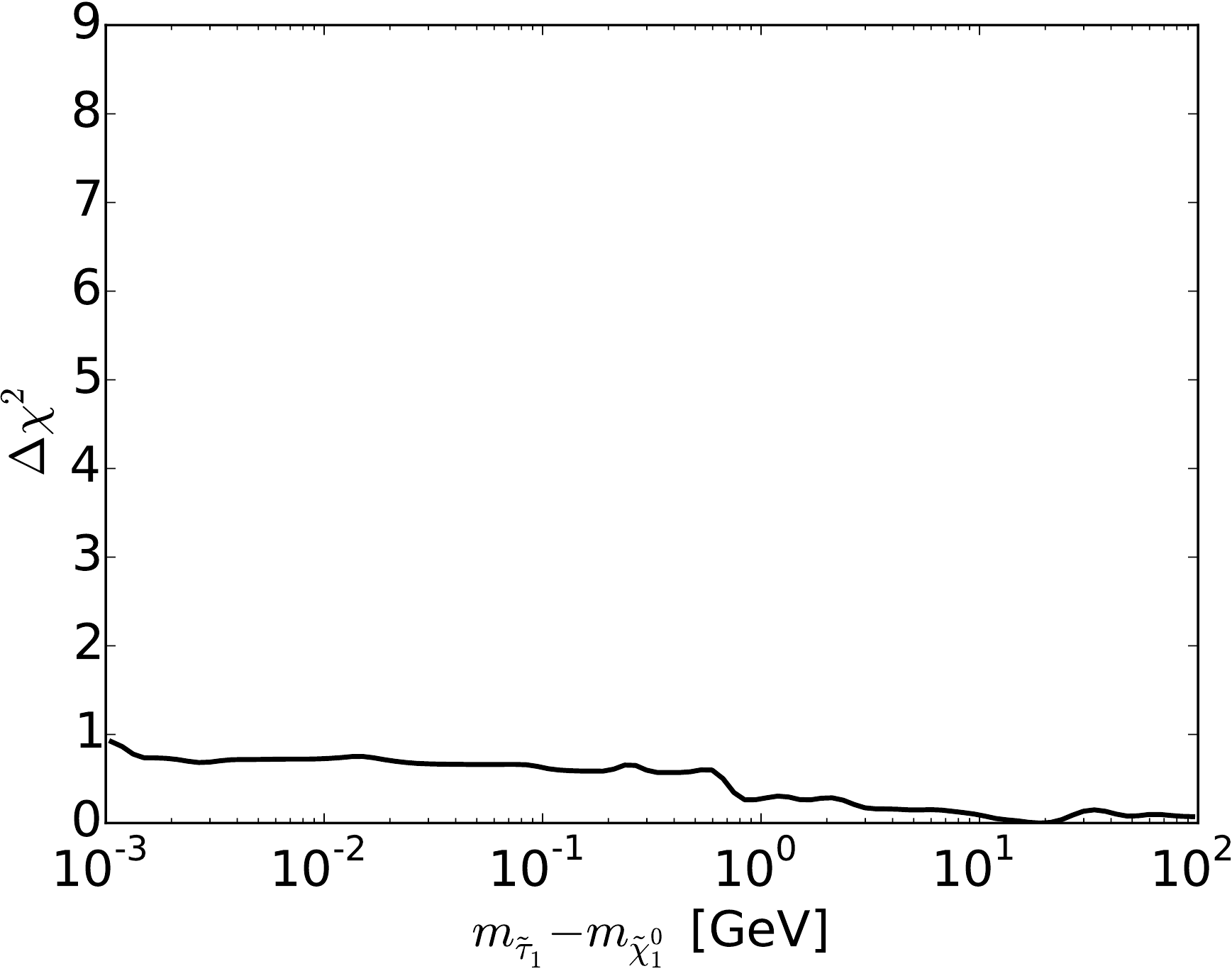}}
\resizebox{7.5cm}{!}{\includegraphics{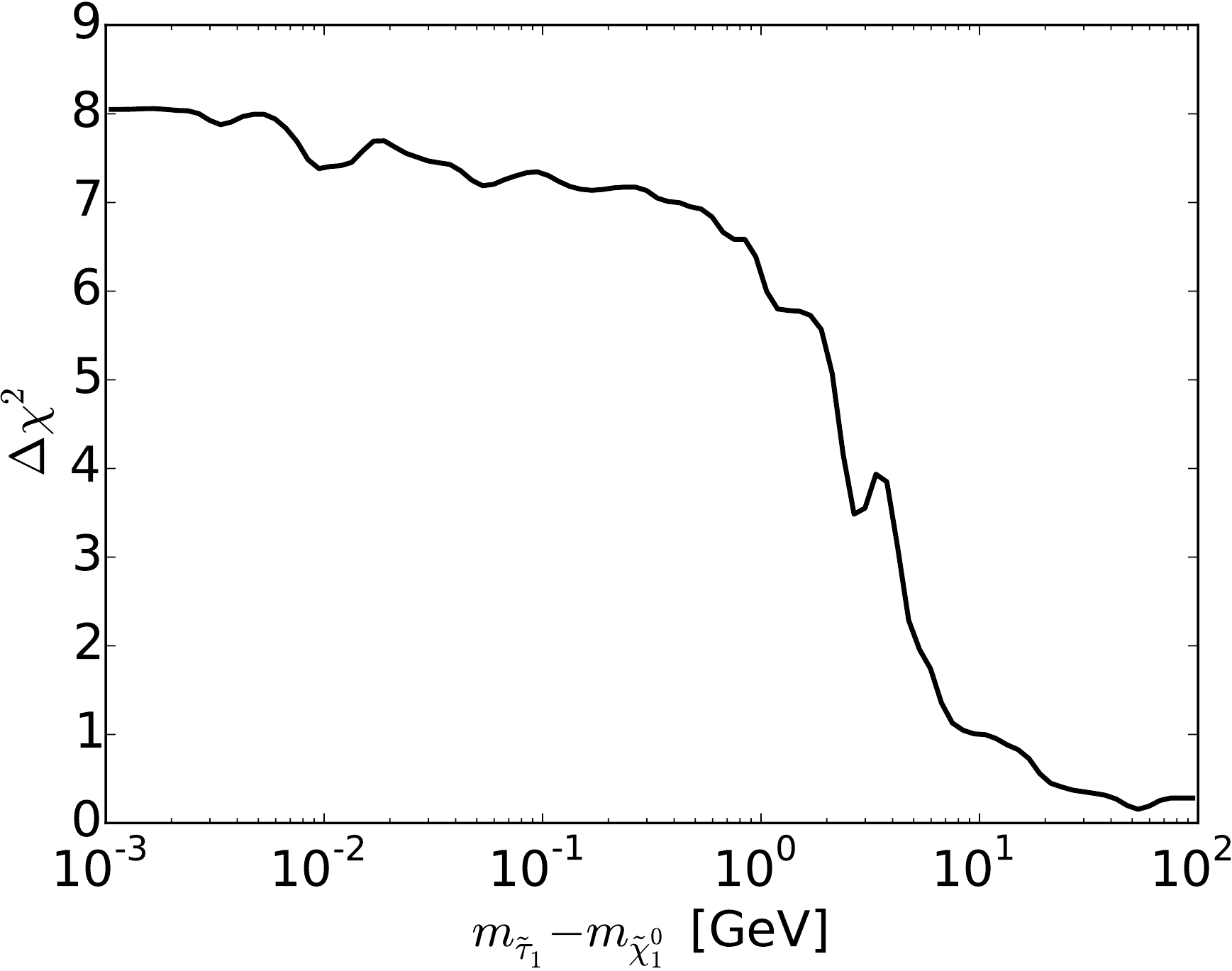}}
\end{center}
\vspace{-0.5cm}
\caption{\it The one-dimensional $\Delta \chi^2$ profile likelihood functions in the CMSSM (upper left),
the NUHM1 (upper right), the NUHM2 (lower left) and the pMSSM10 (lower right) for
$\mstaue - \mneu1 < 100 \gev$. {In the CMSSM, NUHM1 and NUHM2, low values of
$\chi^2$ are found for points with $\mstaue - \mneu1 \sim \mev$,
whereas in the pMSSM10 $\Delta \chi^2$ rises to $\sim 8$ at small $\mstaue - \mneu1$.}
}
\label{fig:stauchi}
\end{figure*}

Fig.~\ref{fig:stauchi} displays on log-linear scales the one-dimensional $\Delta \chi^2$ profile
likelihoods as functions of the ${\staue} - \neu1$ mass differences {in the ${\staue}$ coannihilation
regions in the CMSSM (upper left), the NUHM1 (upper right) and the NUHM2 (lower
left). Also shown (in the lower right panel) is the
corresponding distribution in the pMSSM10, although
in this case the ${\staue}$ coannihilation region
is disfavoured: in this case we see that the likelihood
function increases sharply for ${\mstaue} - \mneu1 \lesssim 10 \gev$, rising to
$\Delta \chi^2 \sim 8$ for very small mass differences.}

In the {CMSSM, NUHM1 and NUHM2} panels the one-dimensional $\chi^2$ profile likelihood function is quite a flat function of
$\mstaue - \mneu1$, {and there are points with $\Delta \chi^2 \lesssim 1$ that have very small
values of this mass difference $\sim \mev$. Hence, it is possible that the $\staue$ might live long enough
($\tau_{\staue} \gtrsim 400$~ns) to appear at the LHC as a long-lived
(LL) charged particle, or even long enough
($\tau_{\staue} \gtrsim 1000$~s) to affect Big Bang nucleosynthesis~\cite{ceflos2,jittoh,Jedamzik}. However, we emphasize that the
mass differences required to realize these possibilities
($\lesssim 1.2 \gev, \lesssim 0.1 \gev$) require quite special
parameter sets}. One should presumably require $\tau_{\staue} \lsim 1000$~s
in order to avoid destroying the success of Big Bang nucleosynthesis,
{a constraint that we impose in the following Figures.}

\begin{figure*}[htb!]
\begin{center}
\resizebox{7.5cm}{!}{\includegraphics{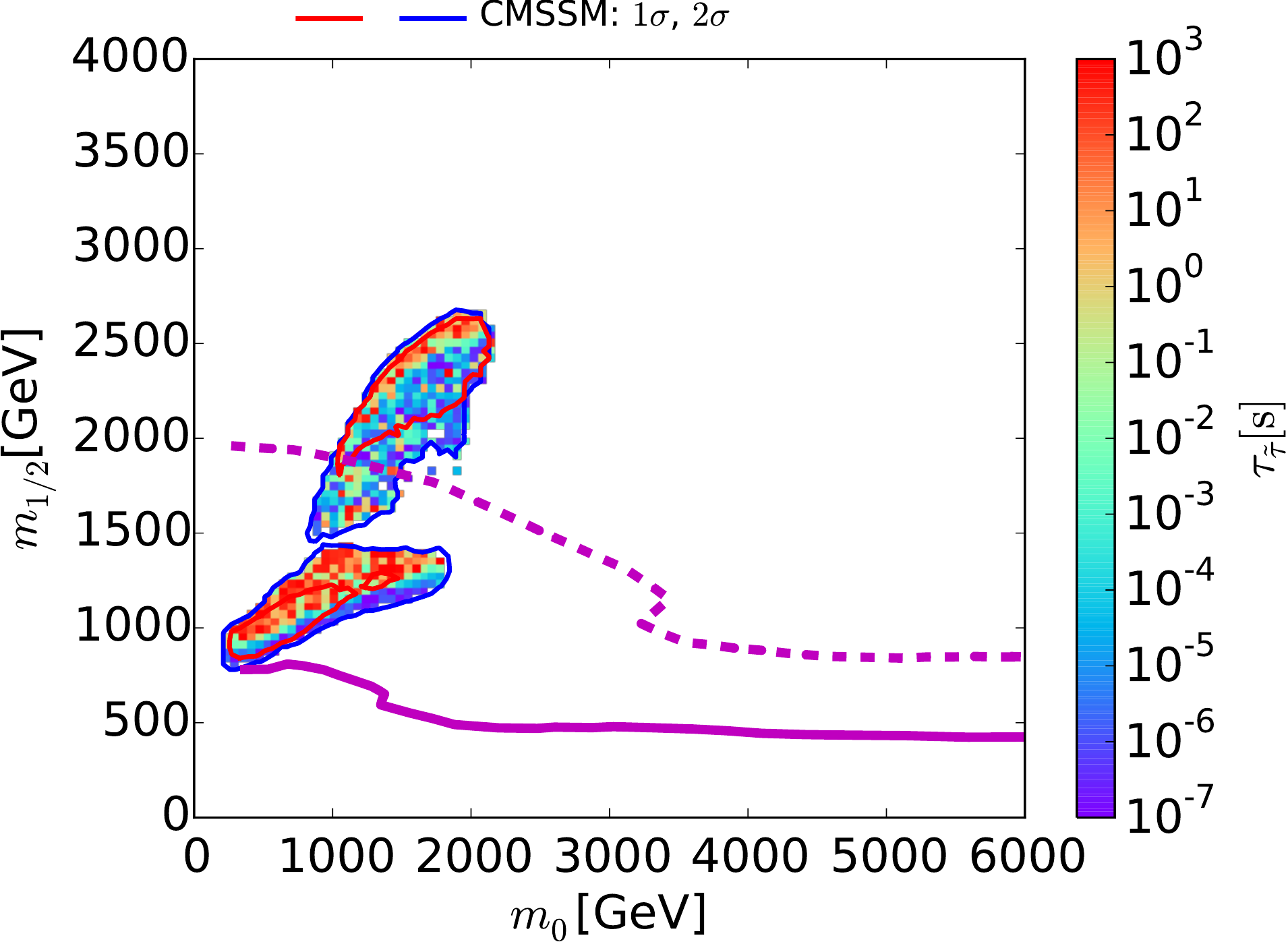}}
\resizebox{7.5cm}{!}{\includegraphics{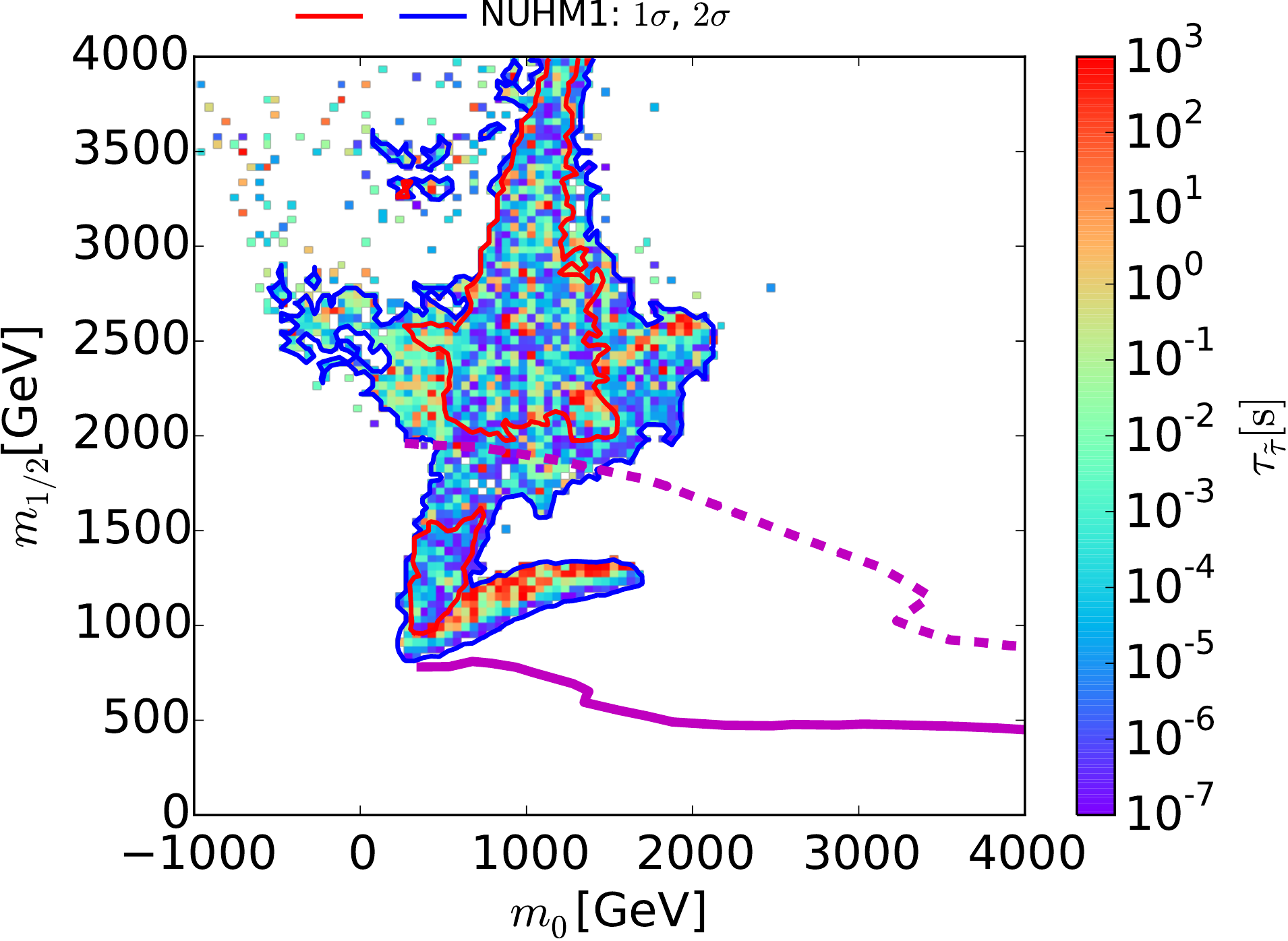}}\\[1em]
\resizebox{7.5cm}{!}{\includegraphics{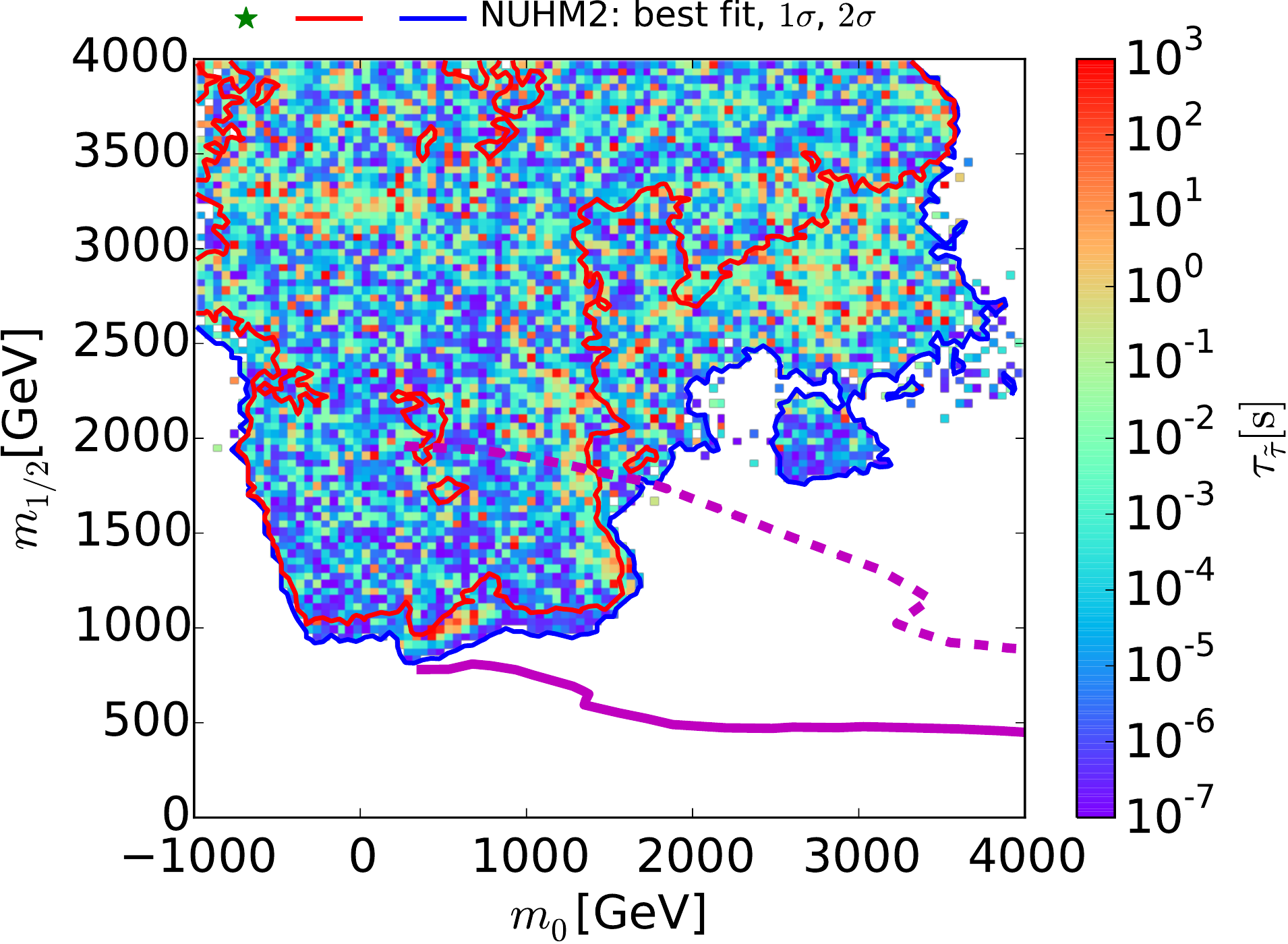}}
\resizebox{7cm}{!}{\includegraphics{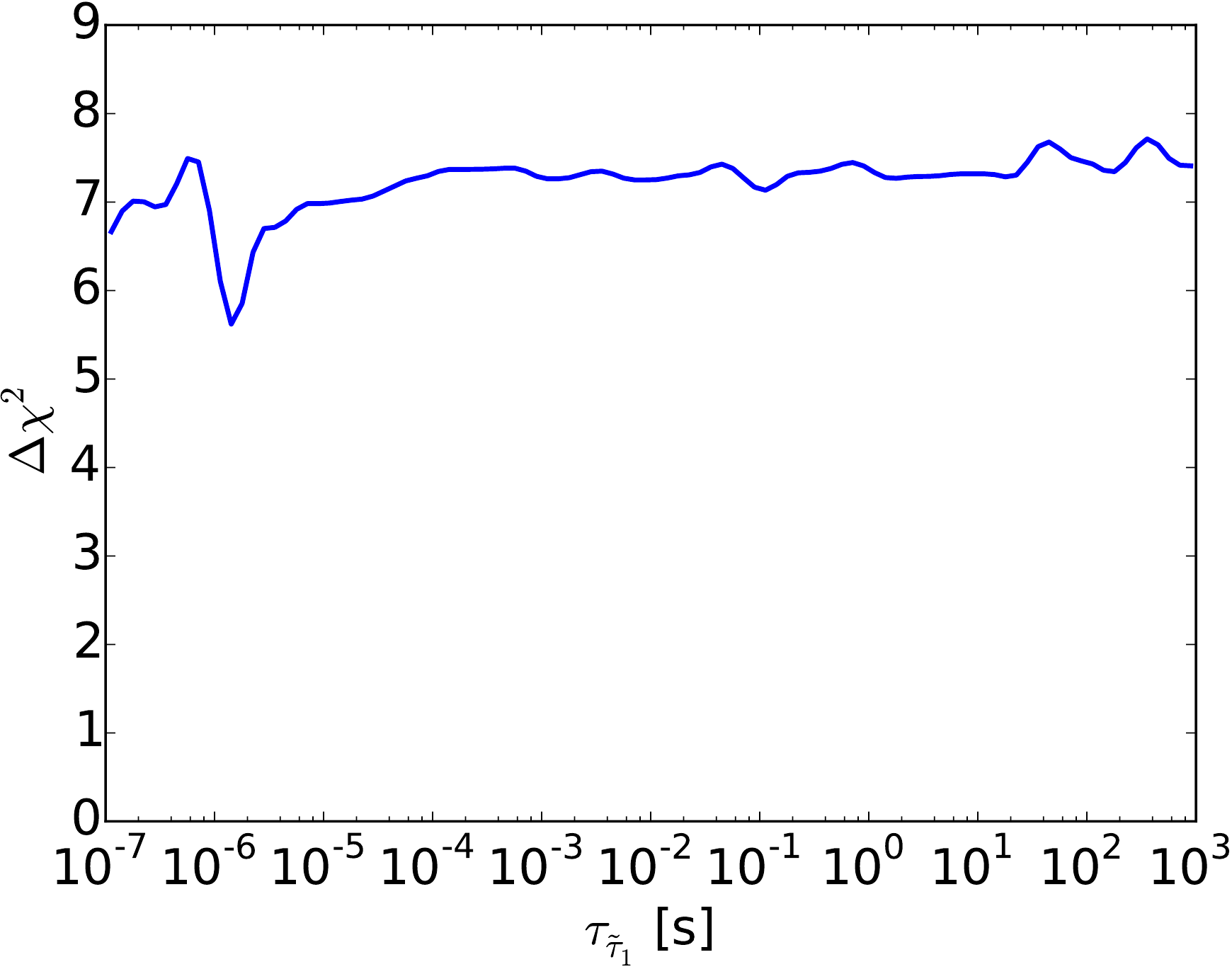}}
\end{center}
\vspace{-0.5cm}
\caption{\it The $(m_0, m_{1/2})$ planes in the CMSSM (upper left), the NUHM1 (upper right) and the NUHM2 (lower left),
showing the regions where {the lowest-$\chi^2$ points have $\Delta \chi^2 < 5.99$ and $10^3$s $> \tau_{\staue} > 10^{-7}$s: 
the lifetimes~\protect\cite{Citron:2012fg} of these points are colour-coded, as indicated in the legends.
The red and blue contours are for $\Delta \chi^2 < 2.30 (5.99)$ relative to the absolute minimum.}
Also shown in these panels as solid purple contours are the current LHC 95\% exclusions from $\ETslash$ searches
in the ${\staue}$ coannihilation regions,
and as dashed purple contours the prospective 5-$\sigma$
discovery reaches for $\ETslash$ searches at the LHC with 3000/fb at 14~TeV, corresponding approximately to the 95\% CL exclusion
sensitivity with 300/fb at 14~TeV. As discussed in the text, the sensitivities of LHC searches for metastable
${\staue}$'s in the ${\staue}$ coannihilation region are expected to be
similar~\protect\cite{Desai:2014uha}. {The lower right panel shows the
one-dimensional $\Delta \chi^2$ function in the pMSSM10 for the lifetime of the ${\staue}$
in the range $10^3$s $> \tau_{\rm NLSP} > 10^{-7}$s.}
}
\label{fig:metastable}
\end{figure*}

If $\mstaue - \mneu1 \lesssim 1.2 \gev$, corresponding to $\tau_{\staue}
\gtrsim 400$~ns, the ${\staue}$ would live long enough to be detectable
at the LHC as a LL charged particle\cite{Desai:2014uha}. 
Fig.~\ref{fig:metastable} displays the regions of the $(m_0, m_{1/2})$ plane in the CMSSM
(upper left panel), the NUHM1 (upper right panel) and the NUHM2 (lower left panel) where
the lowest-$\chi^2$ point has $10^3$s $> \tau_{\staue} > 10^{-7}$s: the lifetimes of these points are colour-coded,
as indicated in the legends.
The contours for $\Delta \chi^2 = 2.30 (5.99)$ relative to the absolute
minimum of our data set
are shown as solid red and blue lines, respectively.

{On the other hand, the lower right panel of Fig.~\ref{fig:metastable} displays the
one-dimensional $\Delta \chi^2$ function in the pMSSM10 for the lifetime of the ${\staue}$
in the range $10^3$s $> \tau_{\staue} > 10^{-7}$s. We see that $\Delta \chi^2 \gtrsim 6$
throughout the displayed range, indicating that a long-lived ${\staue}$ is not expected 
in the region of the pMSSM10 parameter space that is favoured by present data.
This is because in our analysis the \gmt\ measurement and the DM constraint
favour  light sfermions and higgsinos, whereas long-lived charginos typically require 
sfermion and higgsino masses larger than 10 TeV~\cite{1506.08799}~\footnote{{The $\Delta \chi^2$ for a
$\cha{1}$ in our pMSSM10 sample to have a lifetime in the range $10^3$s $> \tau_{\cha{1}} > 10^{-7}$s is much greater,
so this is an even weaker candidate to be a long-lived charged NLSP.}}.}

The sensitivity of the LHC to a long-lived ${\staue}$
has been compared  in~\cite{Desai:2014uha} to the sensitivity to $\ETslash$ events, and found to be comparable within
the uncertainties. We therefore assume that the projected sensitivity of
the LHC to $\ETslash$ events is a good approximation to its sensitivity
to parameter sets in the ${\staue}$ coannihilation region with 
$0.1 \gev < \mstaue - \mneu1 < 1.2 \gev$. {However, while the reach
in the parameter space is similar, the long-lived stau would constitute
a spectacular additional signature, and would give complementary
information to the direct searches for colored sparticles.}
The purple contours in the CMSSM, NUHM1 and NUHM2
panels of Fig.~\ref{fig:metastable} again show the present and prospective reaches of the LHC for such events. We infer that long-lived
charged particles could be detectable at the LHC throughout the lower parts of the allowed regions in the CMSSM,
NUHM1 and NUHM2. {Table~\ref{tab:detectability} also summarizes the observability of
long-lived charged sparticles in the different models considered.}


\subsection{Squark and Gluino Searches}

\begin{figure*}[htb!]
\begin{center}
\resizebox{7.5cm}{!}{\includegraphics{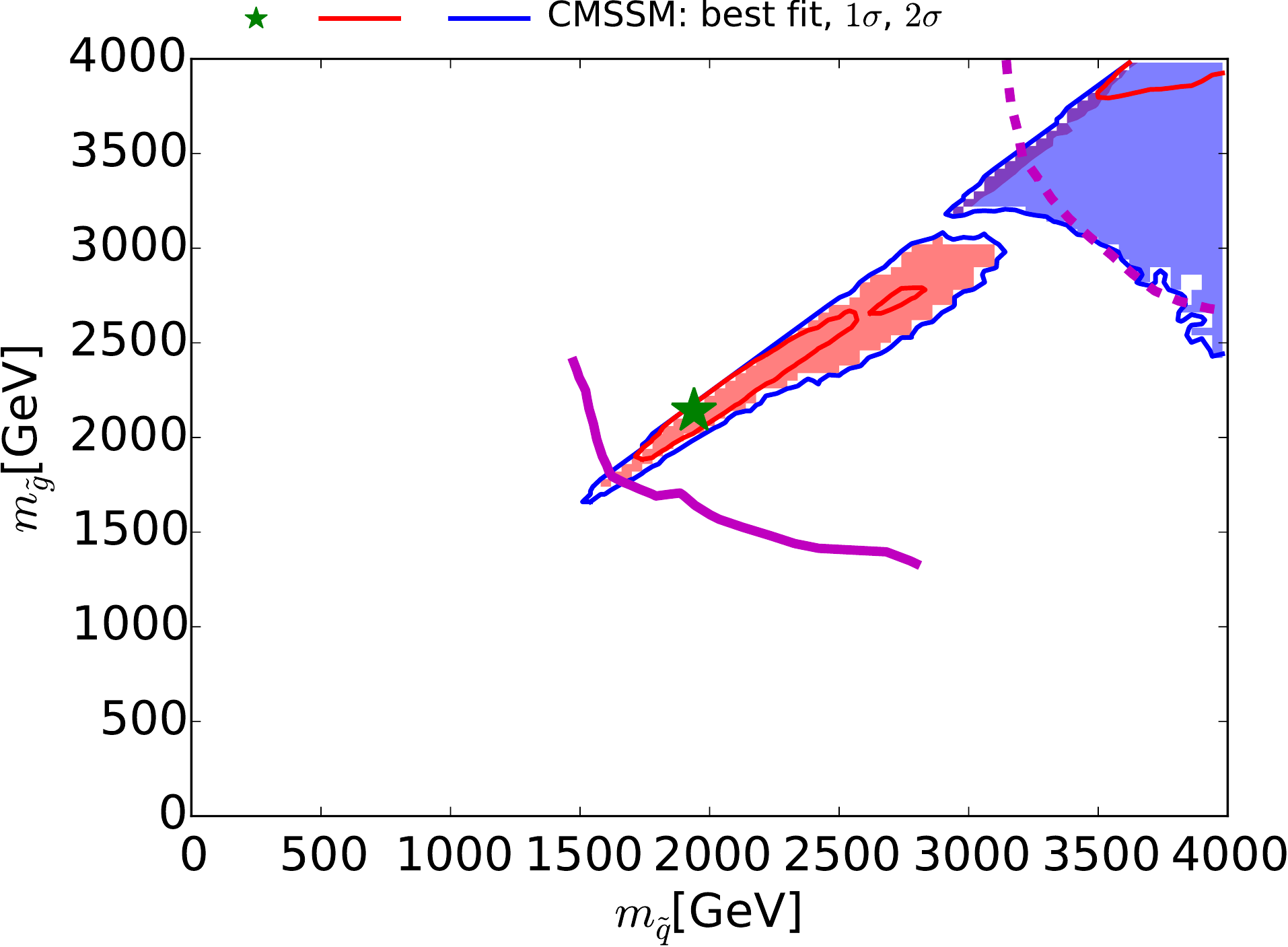}}
\resizebox{7.5cm}{!}{\includegraphics{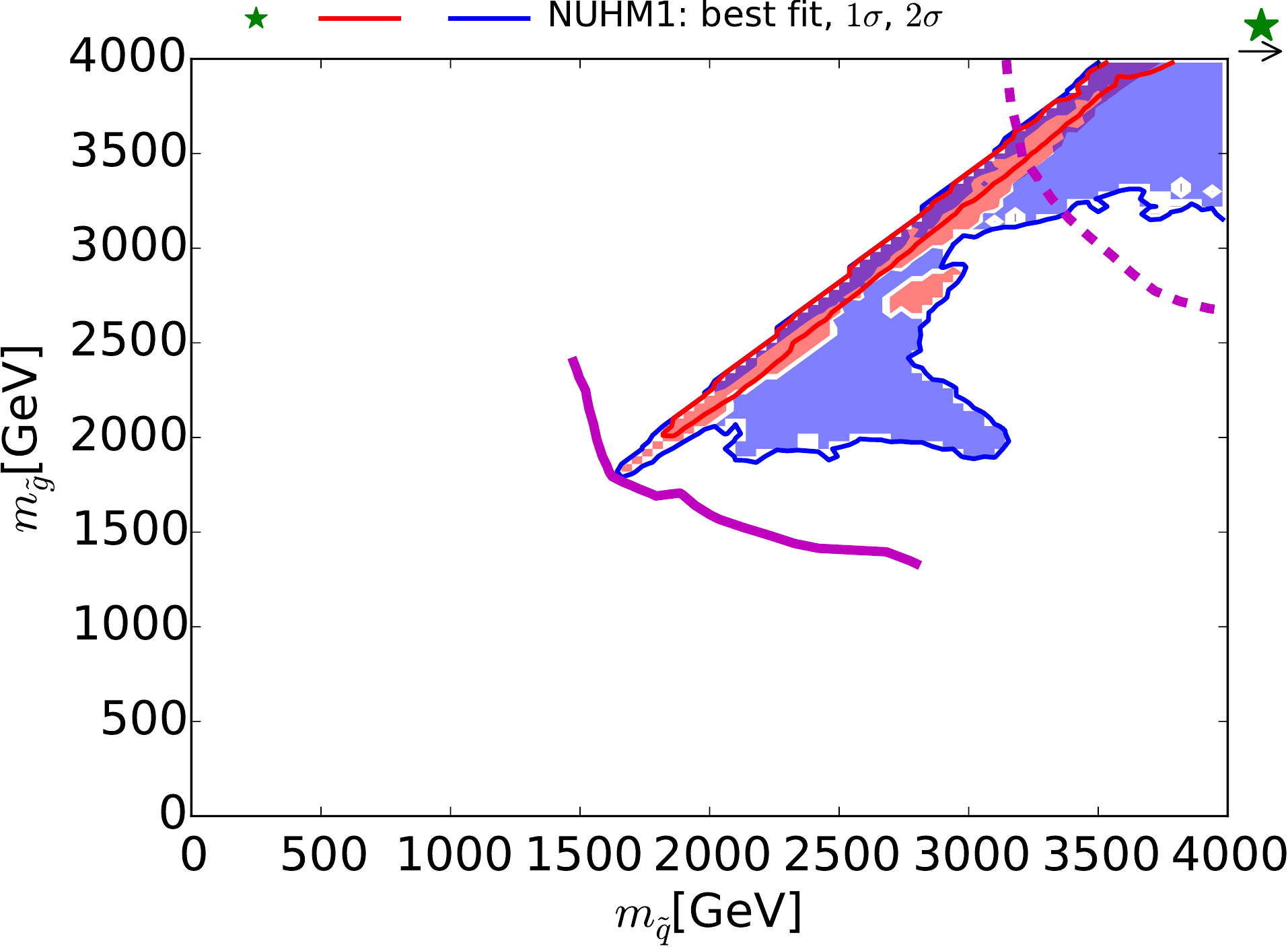}}\\[1em]
\resizebox{7.5cm}{!}{\includegraphics{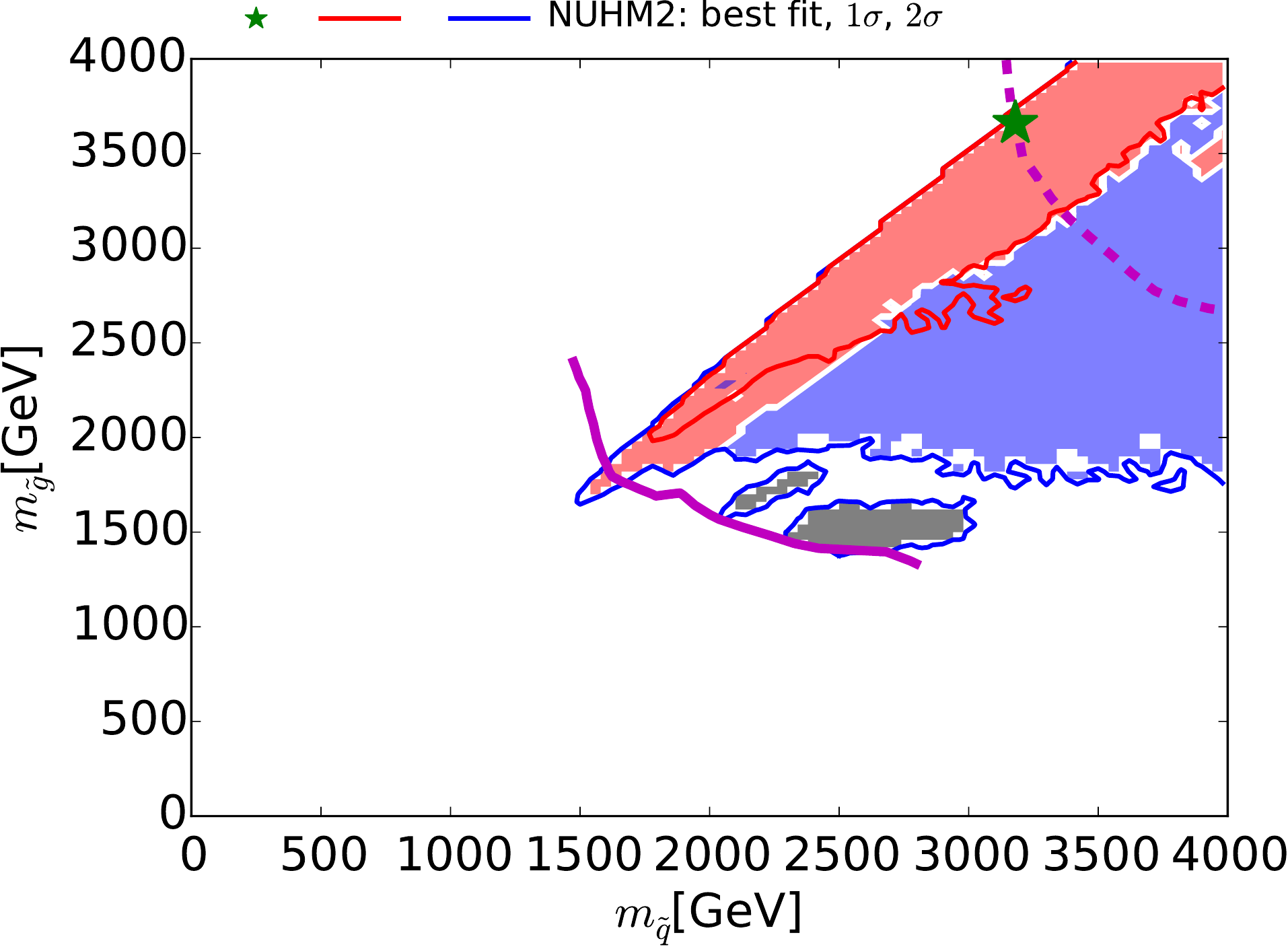}}
\resizebox{7.8cm}{!}{\includegraphics{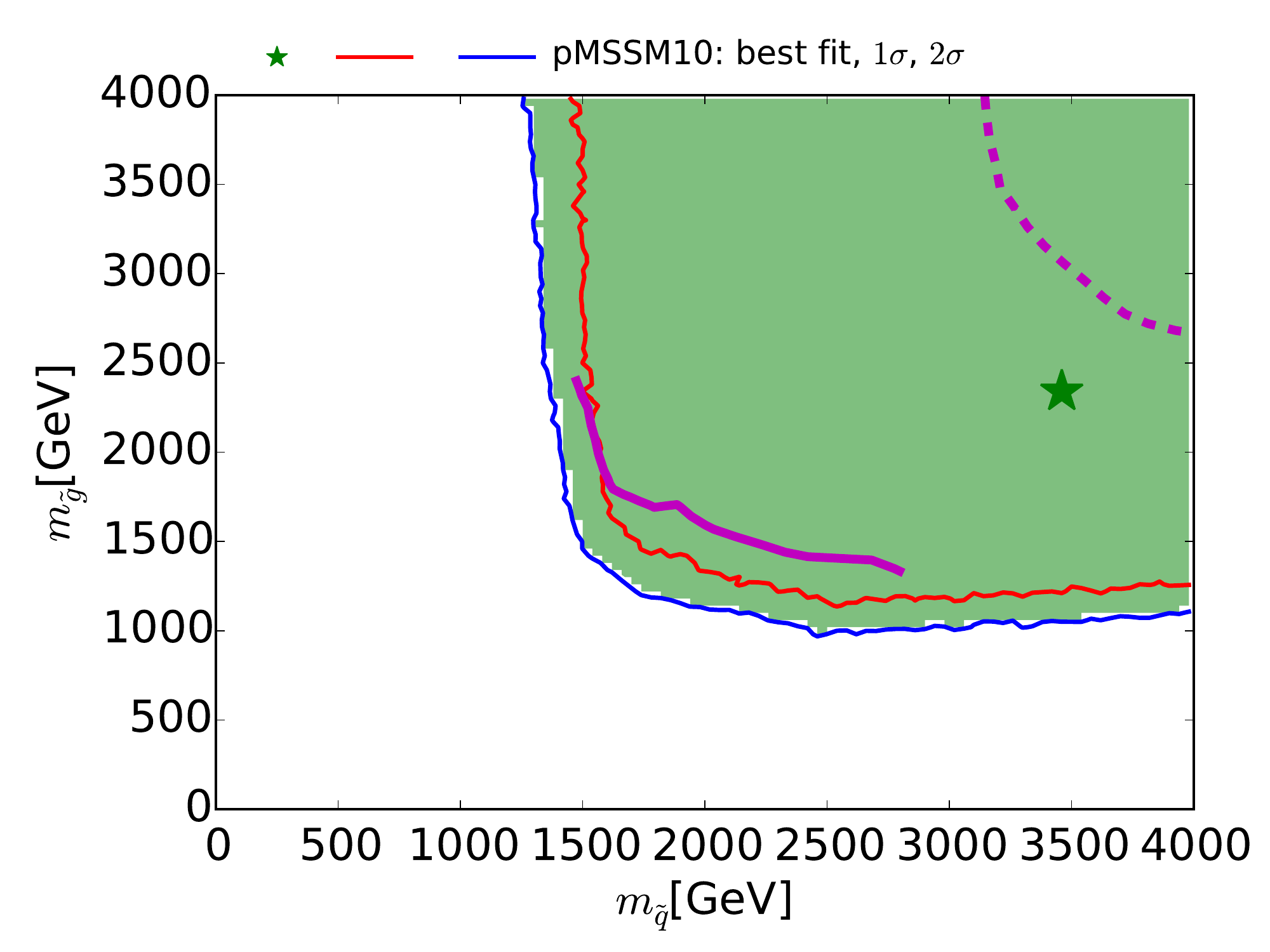}}\\
\resizebox{15cm}{!}{\includegraphics{n12c_dm_legend}}
\end{center}
\vspace{-0.5cm}
\caption{\it The $(\msq, \mgl)$ planes in the CMSSM (upper left),
the NUHM1 (upper right), the NUHM2 (lower left) and the pMSSM10 (lower right).
The red and blue solid lines are  the $\Delta \chi^2 = 2.30$ and 5.99 contours,
and the {solid (dashed) purple lines are the current and (projected) 95\% exclusion contours for ~$\ETslash$ searches
at the LHC (with 300/fb of data at 14~TeV).} The solid lines are almost identical
with the contours for 5-$\sigma$ discovery with 3000/fb.}
\label{fig:msqmgl}
\end{figure*}

Fig.~\ref{fig:msqmgl} displays the $(\msq, \mgl)$ planes for the CMSSM (upper left),
the NUHM1 (upper right), the NUHM2 (lower left) and the pMSSM10 (lower right).
In each panel we show the $\Delta \chi^2 = 2.30$ and 5.99 contours as red and blue
solid lines, respectively. {The current 95\% CL exclusions from ATLAS $\ETslash$ searches
are shown as solid purple lines, and the estimated reaches of $\ETslash$ searches for 95\% exclusion with 300/fb
of data at 14~TeV~\cite{ATLASHiL} (very similar to the reaches for 5-$\sigma$ discovery with 3000/fb)
are shown as dashed purple lines.} The CMSSM panel shows again that
the ${\staue}$ coannihilation region is within the LHC reach in this model. However, in the NUHM1
and the NUHM2 only portions of the ${\staue}$ coannihilation regions are accessible at the LHC,
along with small pieces of the $H/A$ funnel regions. In the case of the NUHM2 the small ${\tilde t_1}$
coannihilation regions are also well within the LHC reach. 

The pMSSM10 panel shows a completely
different picture: $\cha{1}$ coannihilation dominates throughout the
$(\msq, \mgl)$ plane, as discussed at the end of \refse{sec:DDMM}, and the
likelihood function is very flat beyond the current LHC $\ETslash$ exclusion. The LHC at 14~TeV
will explore a large part of the $(\msq, \mgl)$ plane, but a (more) complete exploration would be a
task for a higher-energy collider~\cite{Interplay}.

\subsection{{Charginos and Neutralinos}}

\begin{figure*}[htb!]
\begin{center}
\resizebox{7.5cm}{!}{\includegraphics{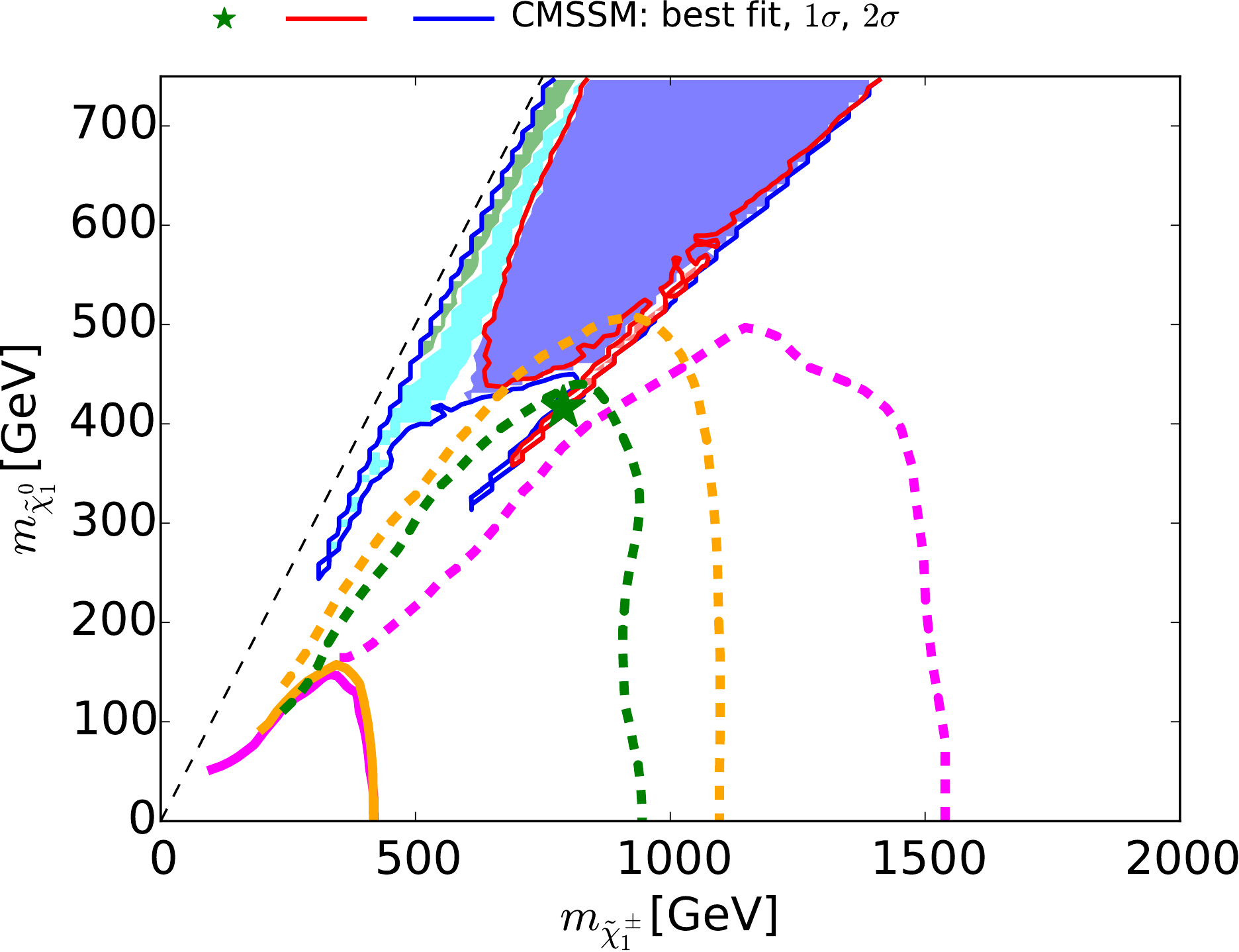}}
\resizebox{7.5cm}{!}{\includegraphics{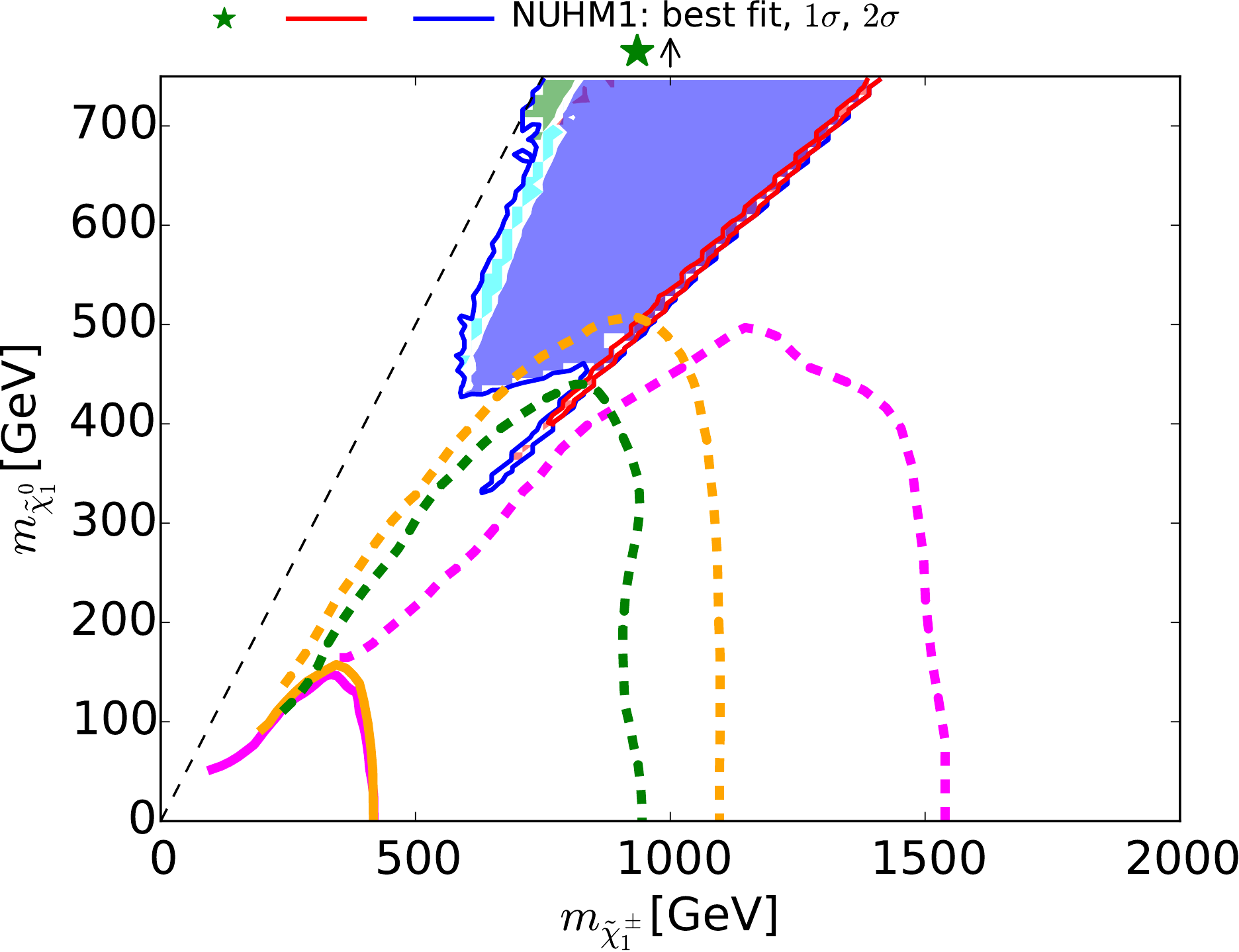}}\\[1em]
\resizebox{7.5cm}{!}{\includegraphics{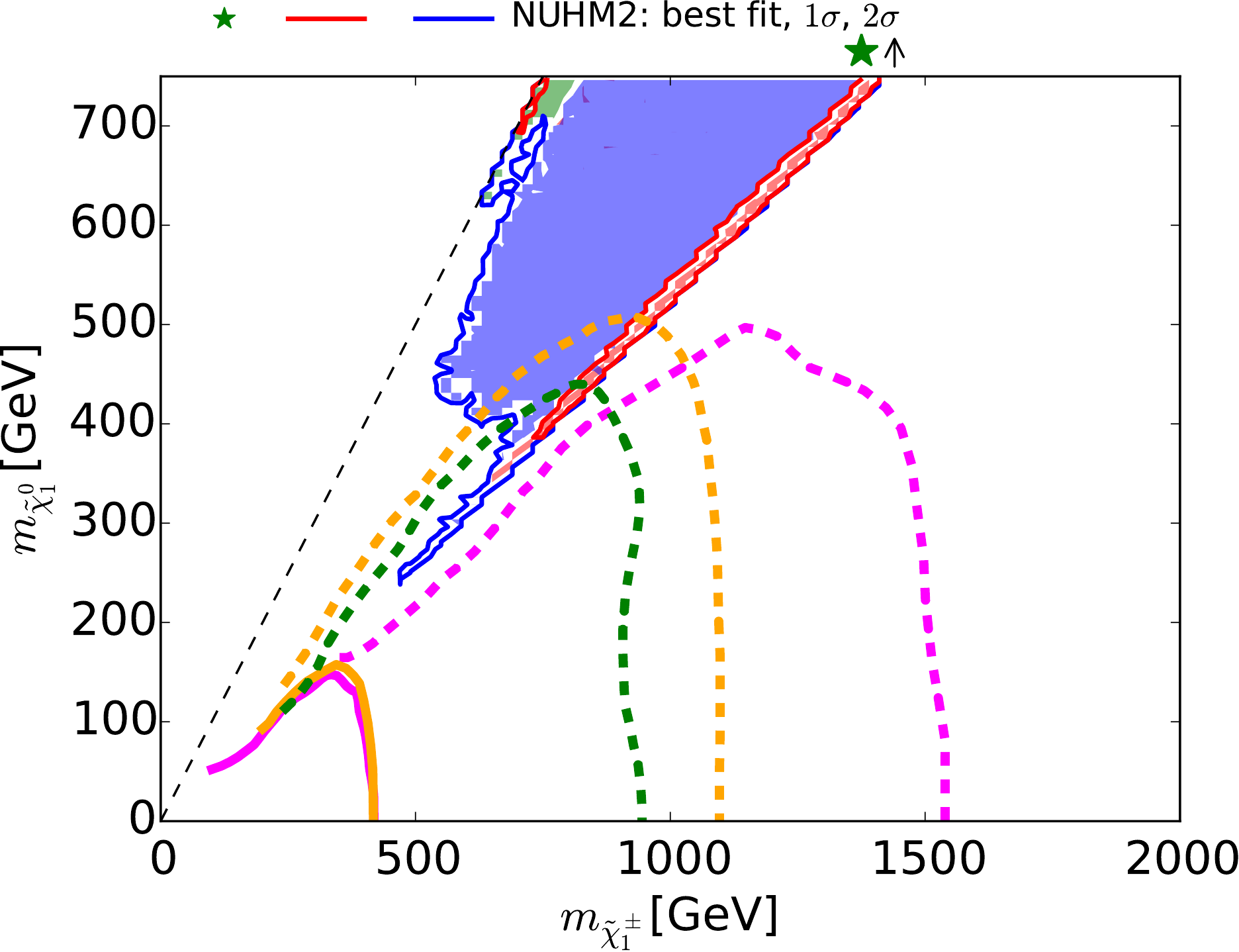}}
\resizebox{7.5cm}{!}{\includegraphics{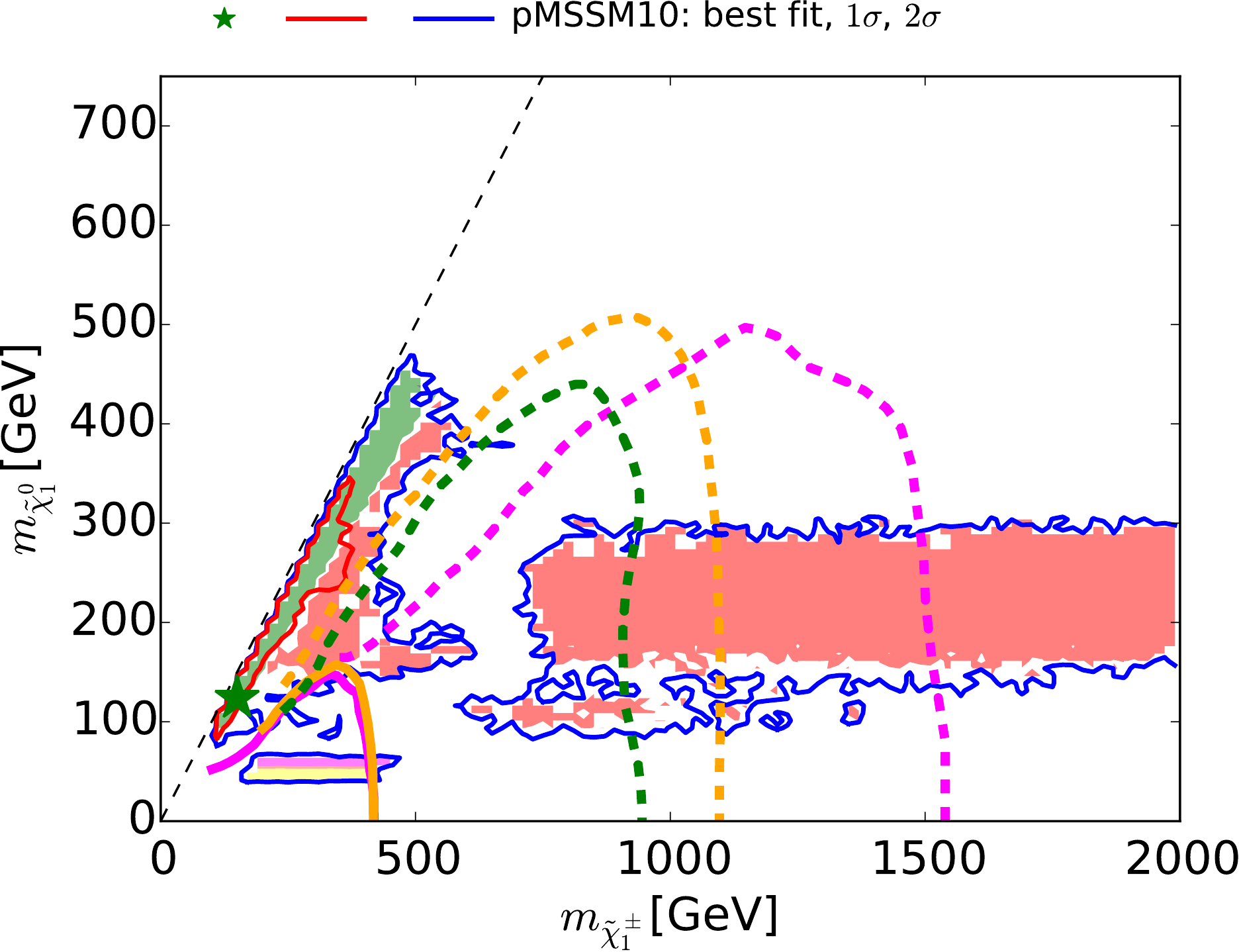}} \\
\resizebox{15cm}{!}{\includegraphics{n12c_dm_legend}}
\end{center}
\vspace{-0.5cm}
\caption{\it The $(\mcha1, \mneu1)$ planes in the CMSSM (upper left),
the NUHM1 (upper right), the NUHM2 (lower left) and the pMSSM10 (lower right).
The red and blue solid lines are  the $\Delta \chi^2 = 2.30$ and 5.99
contours. {The solid (dashed) orange lines are the current and projected 3000/fb 95\% CL exclusion
sensitivities for $\cha1 \neu2 \to W/Z + \ETslash$ searches, the green dashed lines the
projected 3000/fb 95\% CL exclusion sensitivity for a $\cha1 \neu2 \to W/h + \ETslash$ search
(both from~\protect\cite{ATL-PHYS-PUB-2014-010}), and the magenta dashed line is the
projected 3000/fb 95\% CL exclusion sensitivity for $\cha1 \neu2, \cha1 \cha1 \to \tau, {\tilde \tau} \to
2, 3 \tau's + \ETslash$ searches
(from~\protect\cite{1407.0350}).}}
\label{fig:chaneu}
\end{figure*}

The differences between the dominant DM mechanisms in the pMSSM10 and
the other models studied are highlighted in Fig.~\ref{fig:chaneu}, which displays the
$(\mcha1, \mneu1)$ planes in the CMSSM (upper left),
the NUHM1 (upper right), the NUHM2 (lower left) and the pMSSM10 (lower right).
{The diagonal dashed lines indicate where $\mcha1 = \mneu1$.}
As usual, the red and blue solid lines are  the $\Delta \chi^2 = 2.30$ and 5.99 contours.
In the pMSSM10 case, the region preferred at the 68\% CL is a narrow strip where
$\mcha1 - \mneu1$ is small, whereas in the other models much of the 68\% CL region
is in a narrow strip where $\mcha1 \sim 2 \mneu1$. This reflects the fact that in the
CMSSM, NUHM1 and NUHM2 universal boundary conditions are imposed on the 
gaugino masses at the GUT scale. 

{We see that ${\staue}$ coannihilation
dominates over most of the 95\% CL region in this projection of the pMSSM10
parameter space, though not in the 68\% CL region, which
has small $\mcha{1} - \mneu{1}$ and is} where
$\cha{1}$ coannihilation dominates~\footnote{{We also note in Fig.~\ref{fig:chaneu}
the appearance of an uncoloured region with $\mneu{1} \lesssim 150 \gev$, 
which is where `bulk' annihilation dominates.}}.
On the other hand, in the CMSSM, NUHM1 and NUHM2,
the $H/A$ funnel dominates most of the 95\% CL regions in the $(\mcha1, \mneu1)$ planes, and
also the 68\% CL region in the CMSSM, whereas ${\staue}$ coannihilation dominates the 68\%
CL region in the NUHM1 and the NUHM2, as shown in \reffi{fig:m0m12}.
At the 95\% CL there are also small
$\cha{1}$ coannihilation regions in the CMSSM, NUHM1 and NUHM2 where $\mcha1 - \mneu1$ is small,
and the CMSSM and NUHM1 also have small focus-point regions.

Fig.~\ref{fig:chaneu} also displays the present and prospective future sensitivities of
various LHC searches in the $(\mcha1, \mneu1)$ planes. The solid (dashed) orange lines 
are the current and projected 3000/fb 95\% CL exclusion
sensitivities for $\cha1 \neu2 \to W/Z + \ETslash$ searches%
\footnote{{These sensitivities assume that the $\cha1$ and $\neu2$
decay exclusively into the $\neu1$ in association with $W$ and $Z$, 
respectively, not taking into account the decay
$\neu2 \to \neu1 h$~\cite{Bharucha:2013epa,Han:2014nba}.
}}
, the green dashed lines show the
projected 3000/fb 95\% CL exclusion sensitivities for the $\cha1 \neu2 \to W/h + \ETslash$ search,
both taken from~\protect\cite{ATL-PHYS-PUB-2014-010}, and the magenta dashed line is the
projected 3000/fb 95\% CL exclusion sensitivity for $\cha1 \neu2, \cha1 \cha1 \to \tau, {\tilde \tau} \to
2, 3 \tau's + \ETslash$ searches, taken from~\protect\cite{1407.0350}.
We see that these searches have very limited sensitivities to the CMSSM, NUHM1 and NUHM2, but
could explore significant parts of the ${\staue}$ coannihilation region
in the pMSSM10. However, they would 
largely miss the $\cha1$ coannihilation region, 
i.e., the dominant pMSSM10 DM mechanism cannot be explored by direct
searches at the LHC.


{\subsection{The Lighter Stop Squark}}

\begin{figure*}[htb!]
\begin{center}
\resizebox{7.5cm}{!}{\includegraphics{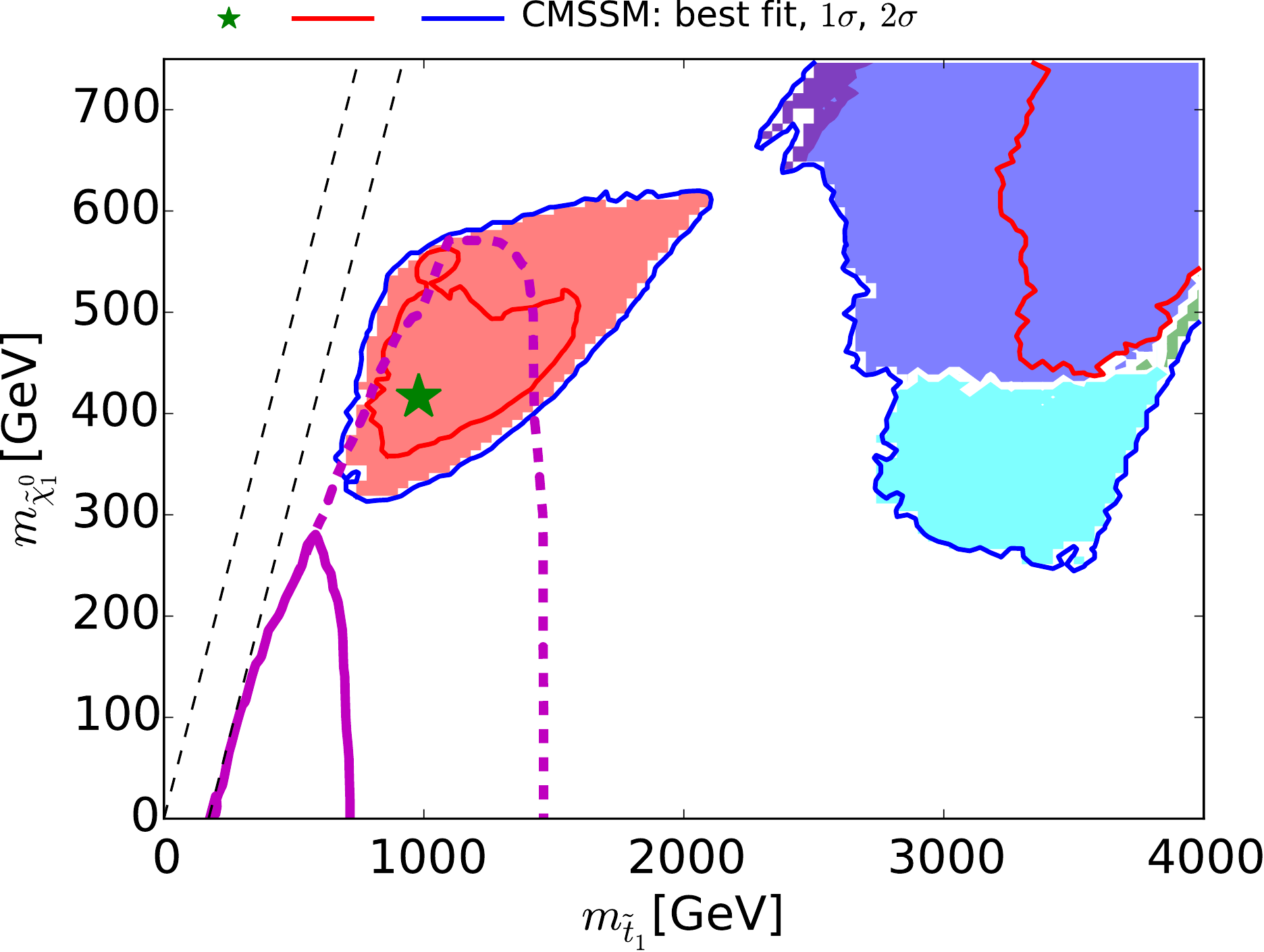}}
\resizebox{7.5cm}{!}{\includegraphics{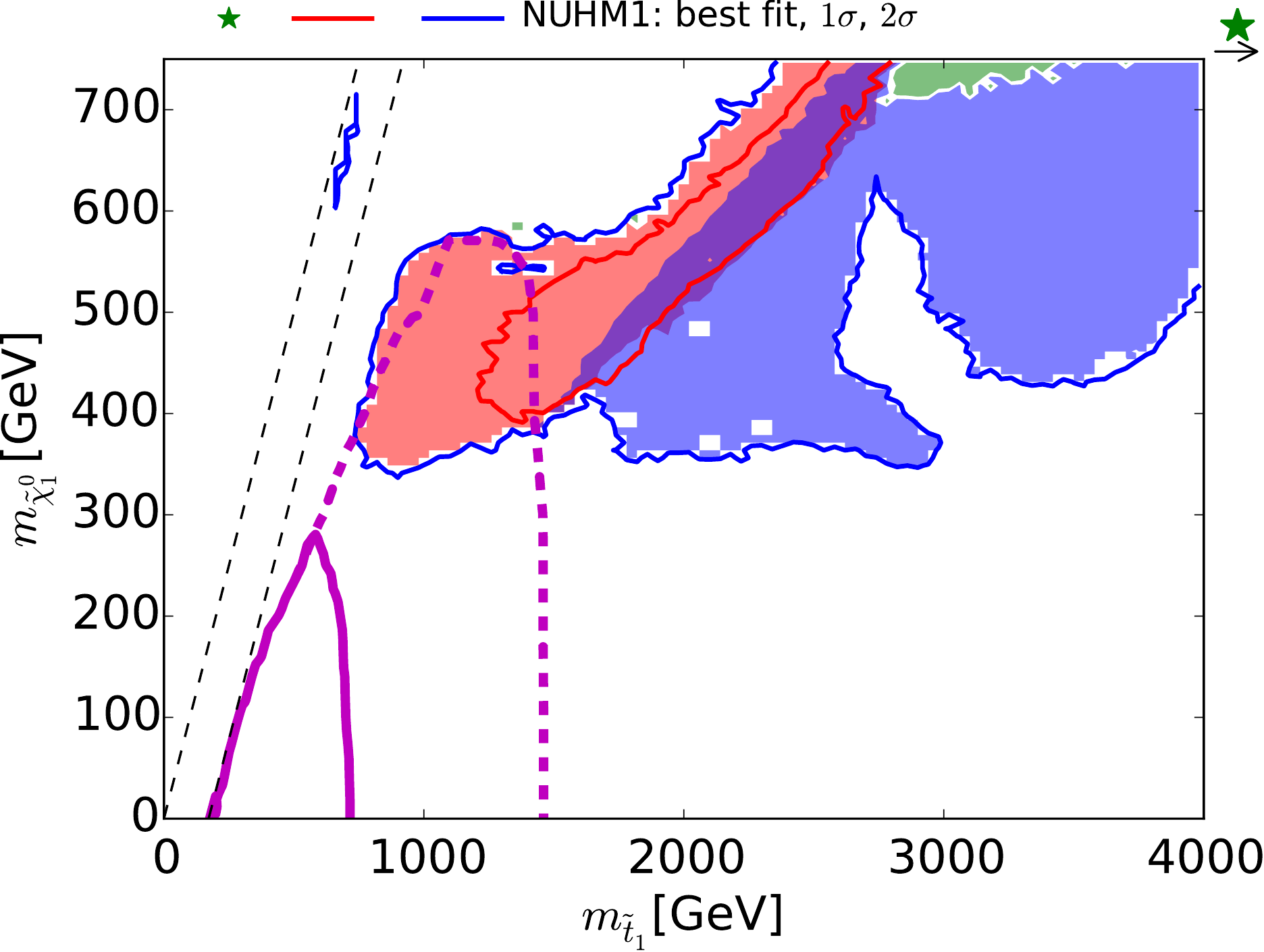}}\\[1em]
\resizebox{7.5cm}{!}{\includegraphics{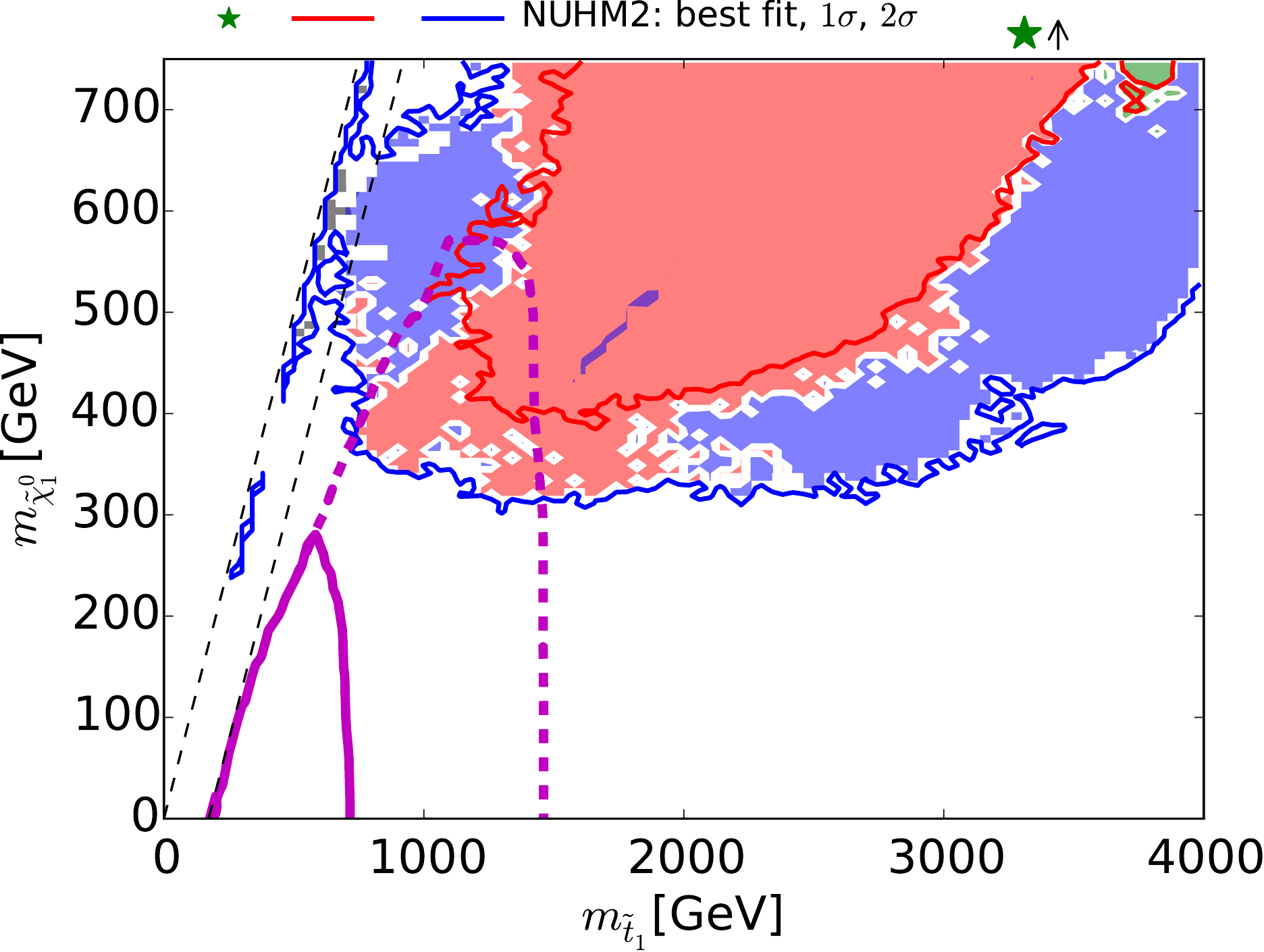}}
\resizebox{7.5cm}{!}{\includegraphics{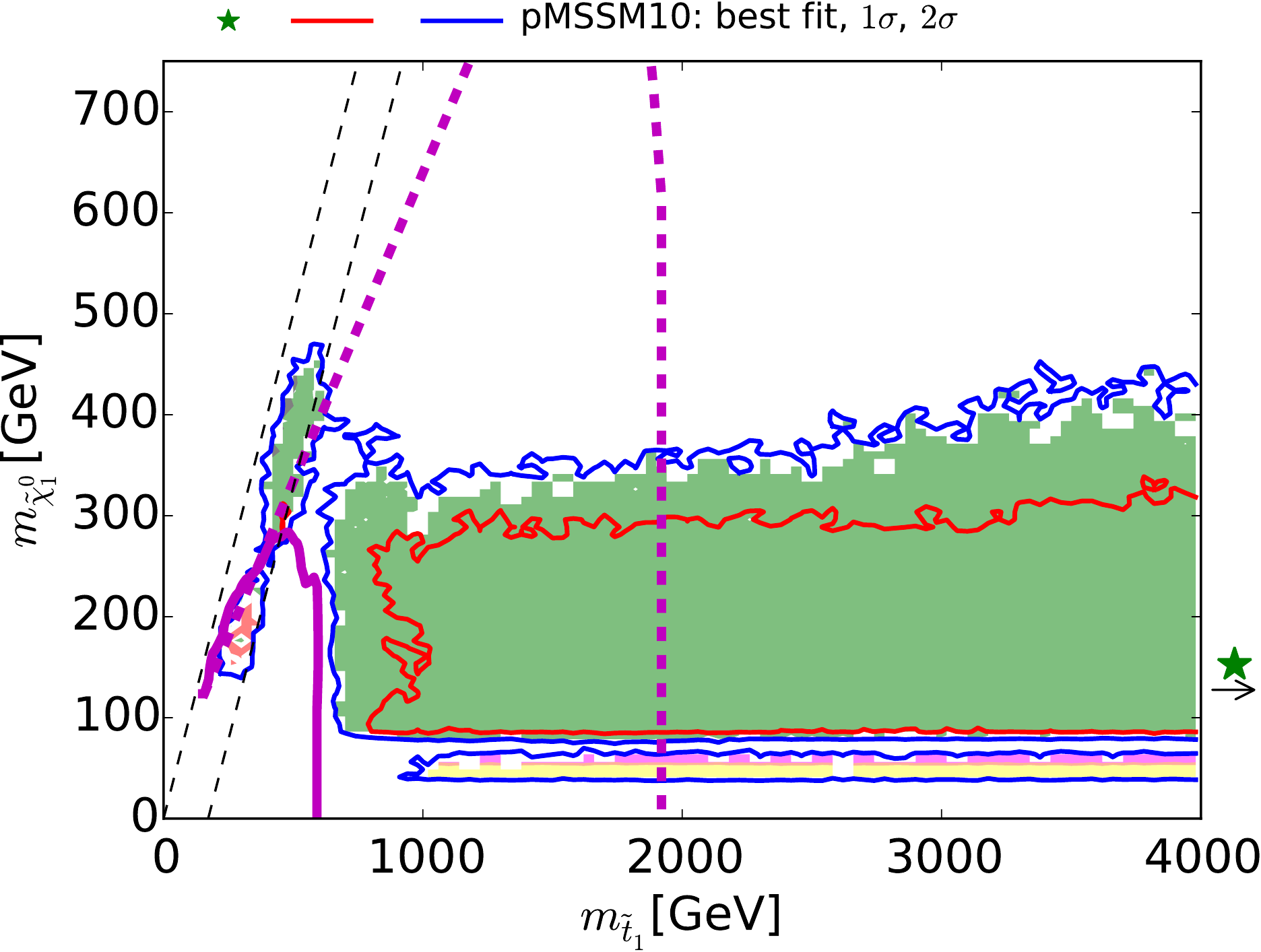}} \\
\resizebox{15cm}{!}{\includegraphics{n12c_dm_legend}}
\end{center}
\vspace{-0.5cm}
\caption{\it The $(m_{\tilde t_1}, \mneu1)$ planes in the CMSSM (upper left),
the NUHM1 (upper right), the NUHM2 (lower left) and the pMSSM10 (lower right).
The red and blue solid lines are  the $\Delta \chi^2 = 2.30$ and $5.99$
contours. The diagonal black dashed lines correspond to $m_{\tilde t_1} = \mneu1$
and $m_{\tilde t_1} = m_t + \mneu1$. In each of the CMSSM, NUHM1 and NUHM2 panels,
the solid purple line is the current 95\% CL limit from the ${\tilde t_1} \to t \neu1$ search in~\protect\cite{1506.08616},
and the dashed purple line is the 3000/fb projection from~\protect\cite{ATL-PHYS-PUB-2013-011}.
{The solid and dashed purple lines for the pMSSM10, obtained using~\protect\cite{1308.2631}
and~\protect\cite{CRtool} respectively, are the LHC Run~1 95\% CL limit and the projected
3000/fb 95\% CL exclusion sensitivity with 3000/fb for a ${\tilde t_1} \to b \cha1$ search,
assuming a 100\% branching ratio and $\mcha1 - \mneu1 = 5 \gev$.}}
\label{fig:stopneu}
\end{figure*}

{Next we study the differences in the impacts of the dominant DM mechanisms on the pMSSM10 and
the other models in the $(m_{\tilde t_1}, \mneu1)$ planes shown in Fig.~\ref{fig:stopneu}. 
{In each of these planes, we indicate by dashed lines where $m_{\tilde t_1} = \mneu1$
and where $m_{\tilde t_1} = m_t + \mneu1$.} In the CMSSM,
as shown in the upper left panel, we see that the ${\staue}$ coannihilation region (which contains all of
the parameter space that is allowed at the 68\% CL) is well separated from
the $H/A$ funnel region, and that only a small part of the displayed portion of the $(m_{\tilde t_1}, \mneu1)$
plane is in the hybrid region. In this model we find that
$m_{\stopone} - \mneu1 \gtrsim 300 \gev$ at the 95\% CL, and we do not find a $\stopone$ coannihilation region}, but we do see
a focus-point region and a small $\cha{1}$ coannihilation region. The situation in the NUHM1 (upper right panel
of Fig.~\ref{fig:stopneu}) exhibits significant differences. The ${\staue}$ coannihilation region
(which again dominates the 68\% CL region) and the $H/A$ funnel
region still dominate the displayed portion of the $(m_{\tilde t_1}, \mneu1)$ plane, but there is a larger hybrid
region, the focus-point region has disappeared and the $\cha{1}$ coannihilation region has remained small,
but has moved to larger $\mneu1$. We also note the appearance of a small $\stopone$ coannihilation `island'
at the 95\% CL in this model. In the case of the NUHM2 (lower left panel), the 68\% CL region is dominated
by ${\staue}$ coannihilation, whereas its extension to the 95\% CL is dominated by the $H/A$ funnel, with
small areas of $\cha{1}$ coannihilation. {In this case there is} a much more
prominent $\stopone$ coannihilation strip at the 95\% CL. Finally, in the pMSSM10 (lower right panel), we see
two {patches with small $m_{\stopone} - \mneu1$ connected by a narrow `isthmus'}, and a `continental' region at large $m_{\stopone}$, where the
68\% CL is located. 
{As indicated by the green and pink shadings,
the dominant DM mechanisms in the `islands' are $\cha1$ and $\staue$ coannihilation, rather than
$\stopone$ coannihilation. We also note the reappearance of the $h$ and $Z$ funnel bands at low $\mneu1$.}
In this model, the lower-mass `island' and part of the higher-mass `island' can be
explored by future LHC searches for ${\stopone} \to b \cha1$~\cite{MC11}, since $\mcha{1} \sim \mneu1$. This search channel is
less powerful in the CMSSM, NUHM1 and NUHM2, where {$\mcha{1} > 2 \mneu1$} in general, particularly in
the ${\staue}$ coannihilation regions that are favoured at the 68\% CL, as seen in Fig.~\ref{fig:chaneu}.

{Fig.~\ref{fig:stopneu} also displays as purple lines the
sensitivities of the {most relevant present (solid) and prospective 3000/fb
(dashed) searches, namely those} for ${\tilde t_1} \to t \neu1$ in the CMSSM, NUHM1
and NUHM2 cases, and for $\sto1 \to b \cha1$ followed by 
$\cha1 \to \neu1$ + soft particles in the pMSSM10 case. 
We see that the current search does not impact the CMSSM, NUHM1 or
NUHM2. In the case of the pMSSM10,  
the solid purple line is the current 95\% CL limit from the
$\sto1 \to b \cha1$ search in~\protect\cite{1308.2631}, assuming a 100\% branching ratio.
This analysis is sensitive to stop topologies with $b$ quarks in the final state where the decay products of the subsequent
$ \cha1$ decay are undetected, and was used in~\cite{MC11} to constrain compressed stop spectra.
However, it is not sensitive to decays involving on-shell $W$ bosons or ${\tilde t_1} \to c \neu1$.
We conclude that future searches have the potential to explore parts of the ${\staue}$ coannihilation regions of
the CMSSM, NUHM1 and NUHM2, and of the $\cha1$ coannihilation region in the pMSSM10 case~\footnote{We
recall that the $h$ and $Z$ funnels in the pMSSM10 could in principle be explored by future searches
for invisible $h$ and $Z$ decays.}}{, but no DM channel can be fully
explored by LHC searches.}

~~\\
\subsection{The Heavy Higgs Bosons}

\begin{figure*}[htb!]
\begin{center}
\resizebox{7.5cm}{!}{\includegraphics{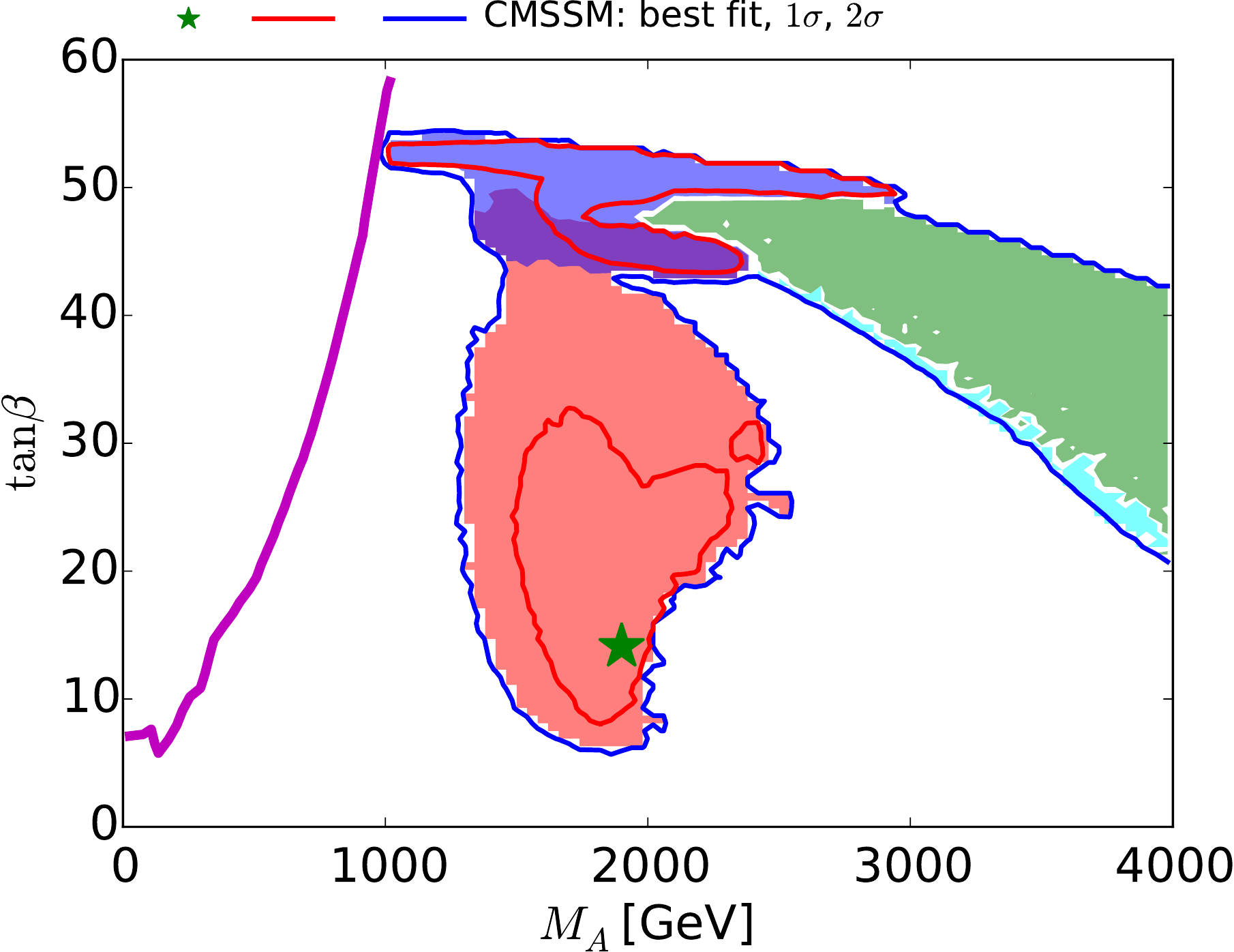}}
\resizebox{7.5cm}{!}{\includegraphics{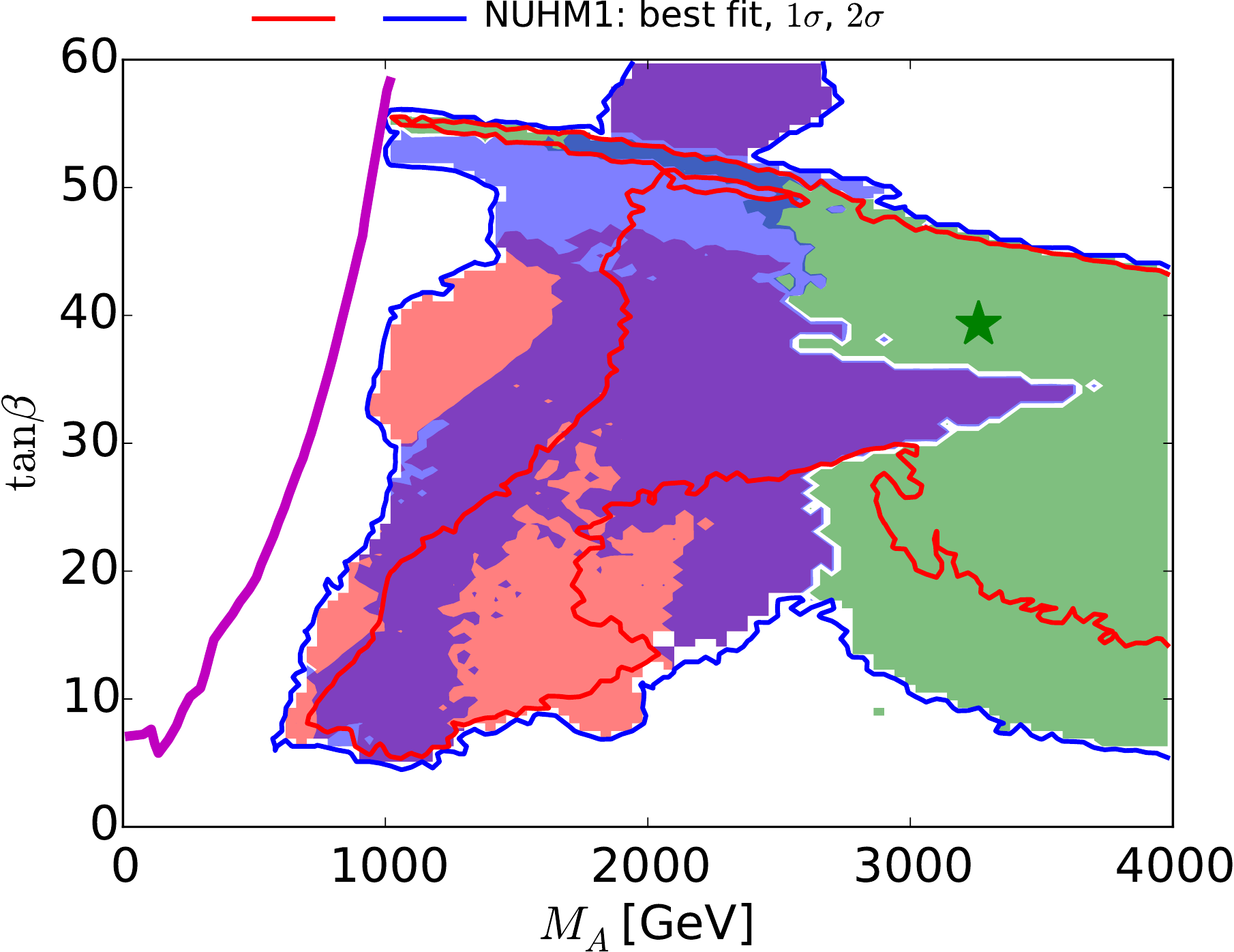}}\\[1em]
\resizebox{7.5cm}{!}{\includegraphics{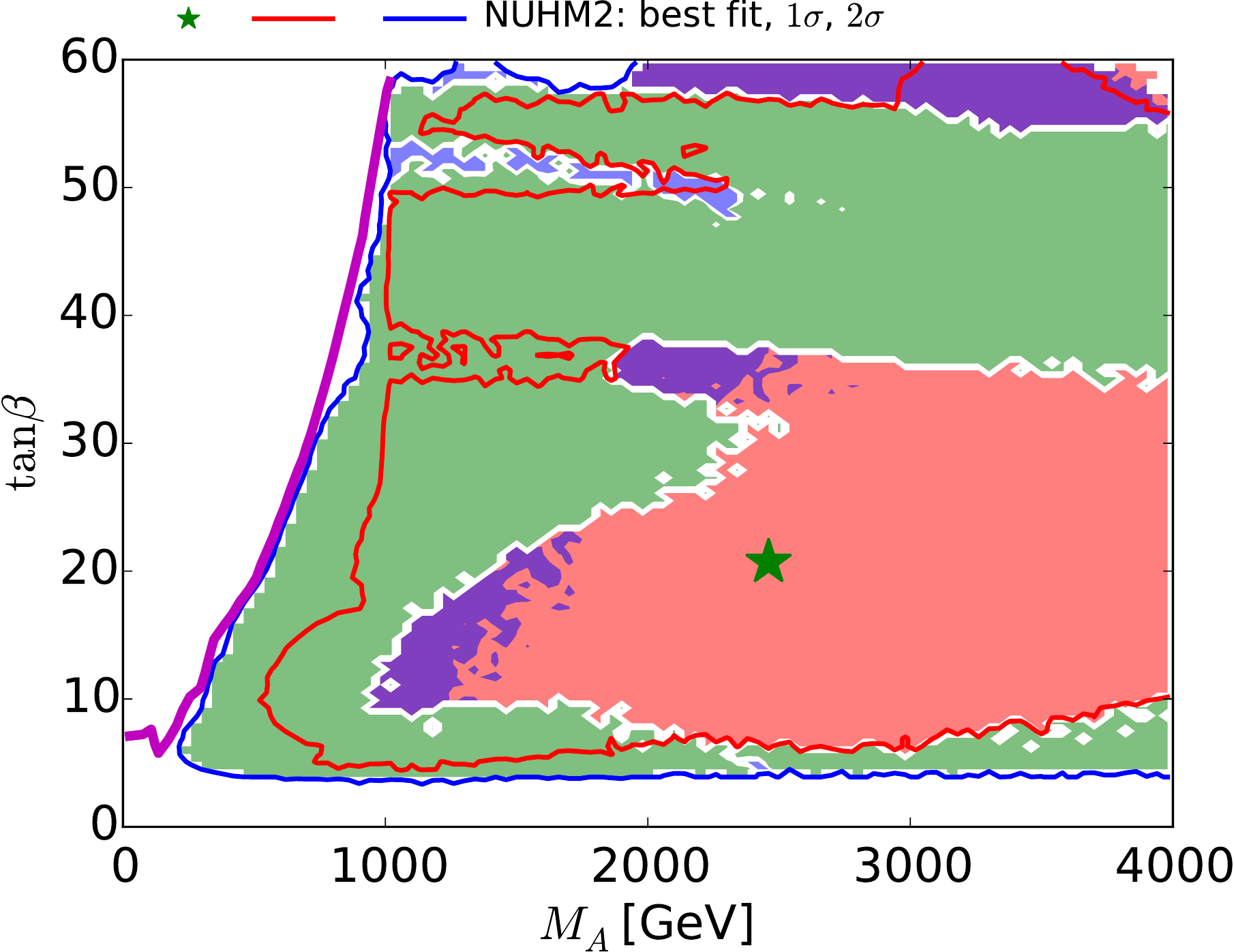}}
\resizebox{7.5cm}{!}{\includegraphics{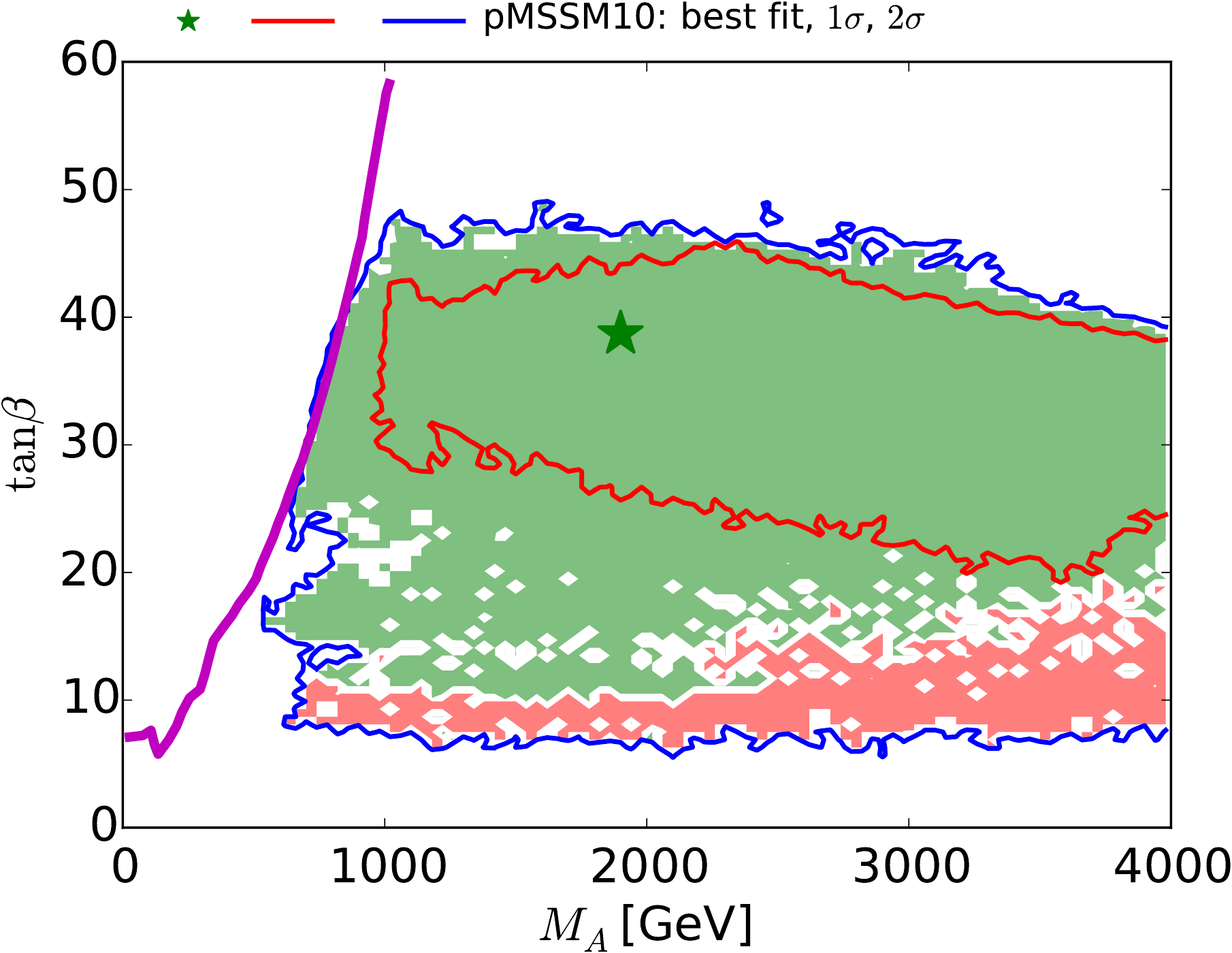}}\\
\resizebox{15cm}{!}{\includegraphics{n12c_dm_legend}}
\end{center}
\vspace{-0.5cm}
\caption{\it The $(\MA, \tb)$ planes in the CMSSM (upper left),
the NUHM1 (upper right), the NUHM2 (lower left) and the pMSSM10 (lower right).
The red and blue solid lines are  the $\Delta \chi^2 = 2.30$ and 5.99
contours, {and the solid purple line is the current LHC 95\% CL exclusion in the $\Mh^{\rm max}$
scenario.}}
\label{fig:MAtb}
\end{figure*}

We now study the differences between the dominant DM mechanisms in the pMSSM10 and
the other models in the $(\MA, \tb)$ planes shown in Fig.~\ref{fig:MAtb}. In the case of the CMSSM, the
regions allowed at the 95\% CL and preferred at the 68\% CL (blue and red contours, respectively)
are generally at considerably larger $\MA$ than the LHC bound (shown as
a solid purple line)~\cite{ATLASHA}~\footnote{This line was calculated assuming the specific MSSM $m_h^{\rm max}$
scenario with $M_{\rm SUSY} = 1 \tev$~\cite{mhmax}, and is used to
{give a rough impression of the location of the direct heavy
  Higgs-boson search bounds}.}.
The ${\staue}$ coannihilation mechanism dominates in a region around $\MA \sim 2000 \gev$
for $\tb \lesssim 40$, the $H/A$ funnel dominates for $\tb \sim 50$, and there is an intermediate hybrid region.
On the other hand, $\cha{1}$ coannihilation dominates in an arc at larger $\MA$.
In the NUHM1, the hybrid and $\cha{1}$ coannihilation regions are greatly expanded, and values of
$\MA$ closer to the LHC bound are allowed at the 95\% CL. In the NUHM2, on the other hand,
essentially all values of $\MA$ consistent with the LHC bound are allowed at the 95\% CL,
the $\cha{1}$ coannihilation mechanism dominates over most of the $(\MA, \tb)$ plane, leaving
${\staue}$ coannihilation to dominate for $\MA \gtrsim 2000 \gev$  and $\tb \lesssim 30$.
Finally, we see that in the pMSSM10 $\cha{1}$ coannihilation dominates the 68\% CL region,
that there is also a region at $\tb \lesssim 20$ where ${\staue}$ coannihilation may be important,
{and that there are intermediate uncoloured regions where neither of these mechanisms dominate}.
The LHC bound on $\MA$ is again saturated for $\tb \lesssim 50$.

We have also estimated (not shown) the prospective LHC 95\% CL
exclusion sensitivity in the $H/A$ plane with 300/fb of data for the
$m_h^{\rm max}$ 
scenario, scaling the current limit {(using results
from~\cite{FeynHiggs} and~\cite{SusHi}), and comparing with the
estimated limits in~\cite{Djouadietal}, where good overall agreement was
found.} We estimate that $\MA \lesssim 2 \tev$ could be explored for
$\tb \sim 50$, 
reducing to $\MA \lesssim 1 \tev$ for $\tb \sim 20$. This would cover much of the $H/A$ funnel
and hybrid regions in the CMSSM, portions of these regions in the NUHM1, and parts of the
$\cha1$ coannihilation regions in the NUHM2 and the pMSSM10,
though not the regions of ${\staue}$ coannihilation in these models. {Table~\ref{tab:detectability} also summarizes the observability of
the heavy Higgs bosons $A/H$ in the different scenarios considered.}


\section{Direct Dark Matter Searches}

\begin{figure*}[htb!]
\begin{center}
\resizebox{7.5cm}{!}{\includegraphics{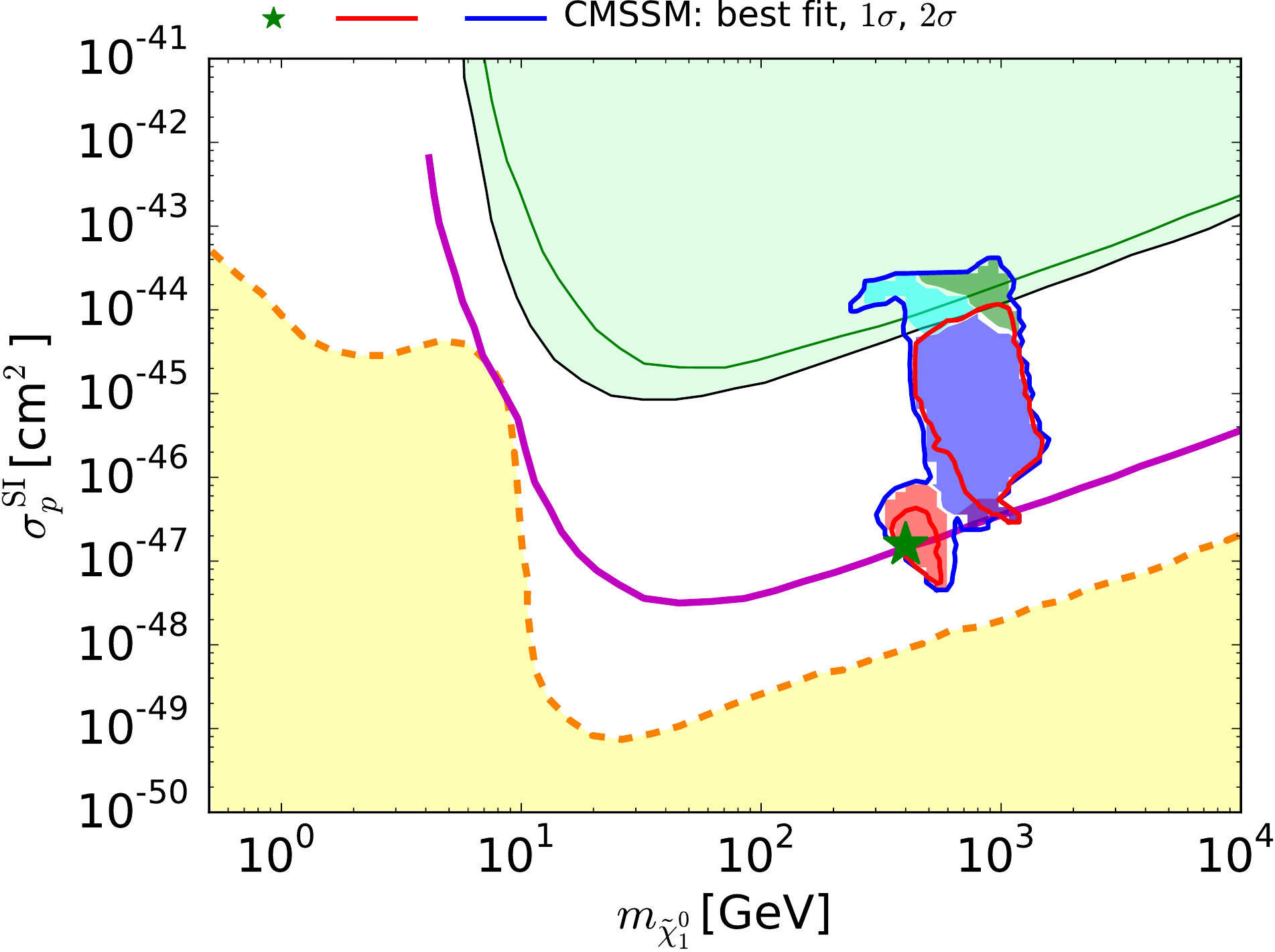}}
\resizebox{7.5cm}{!}{\includegraphics{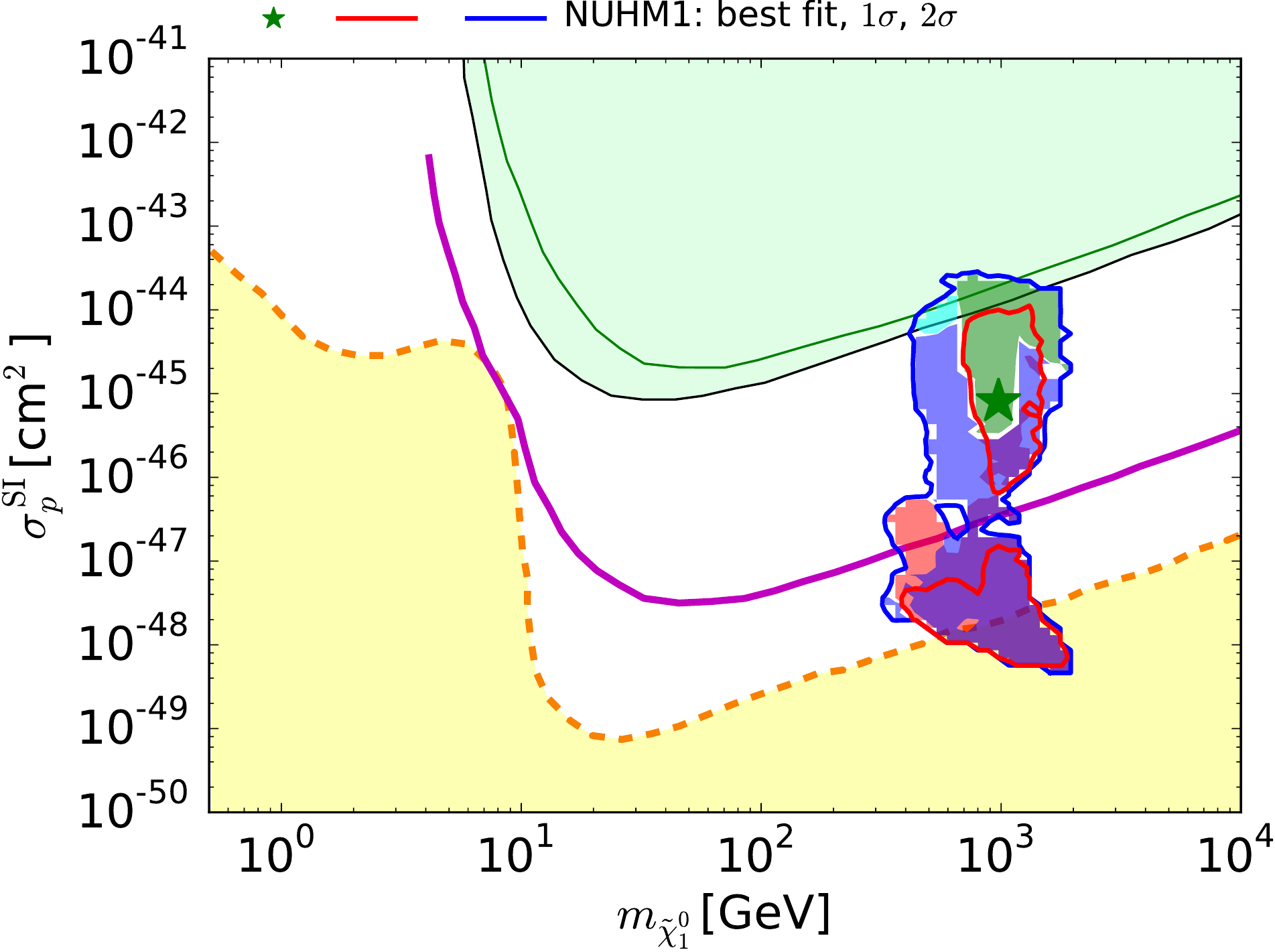}}\\[1em]
\resizebox{7.5cm}{!}{\includegraphics{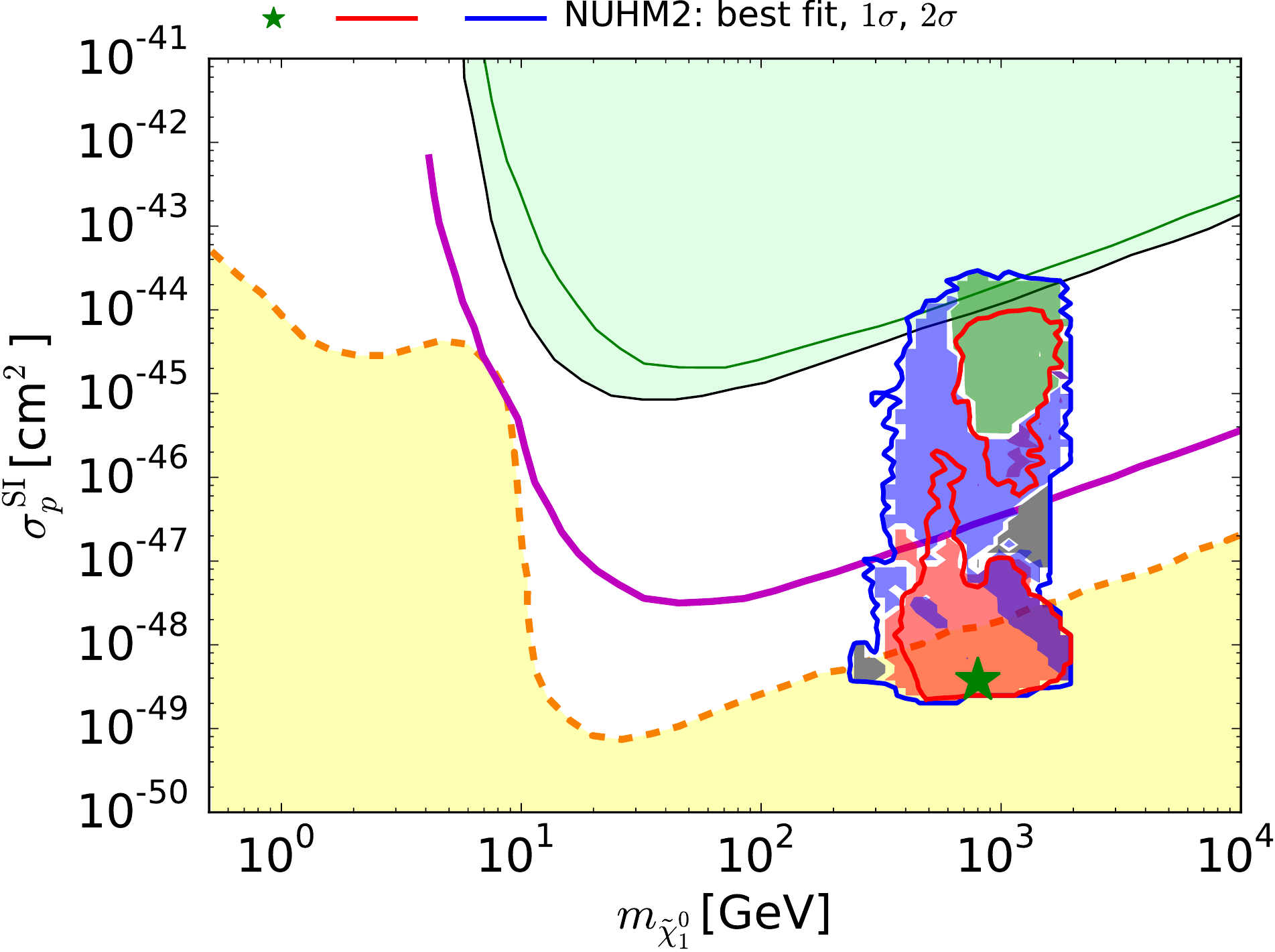}}
\resizebox{7.5cm}{!}{\includegraphics{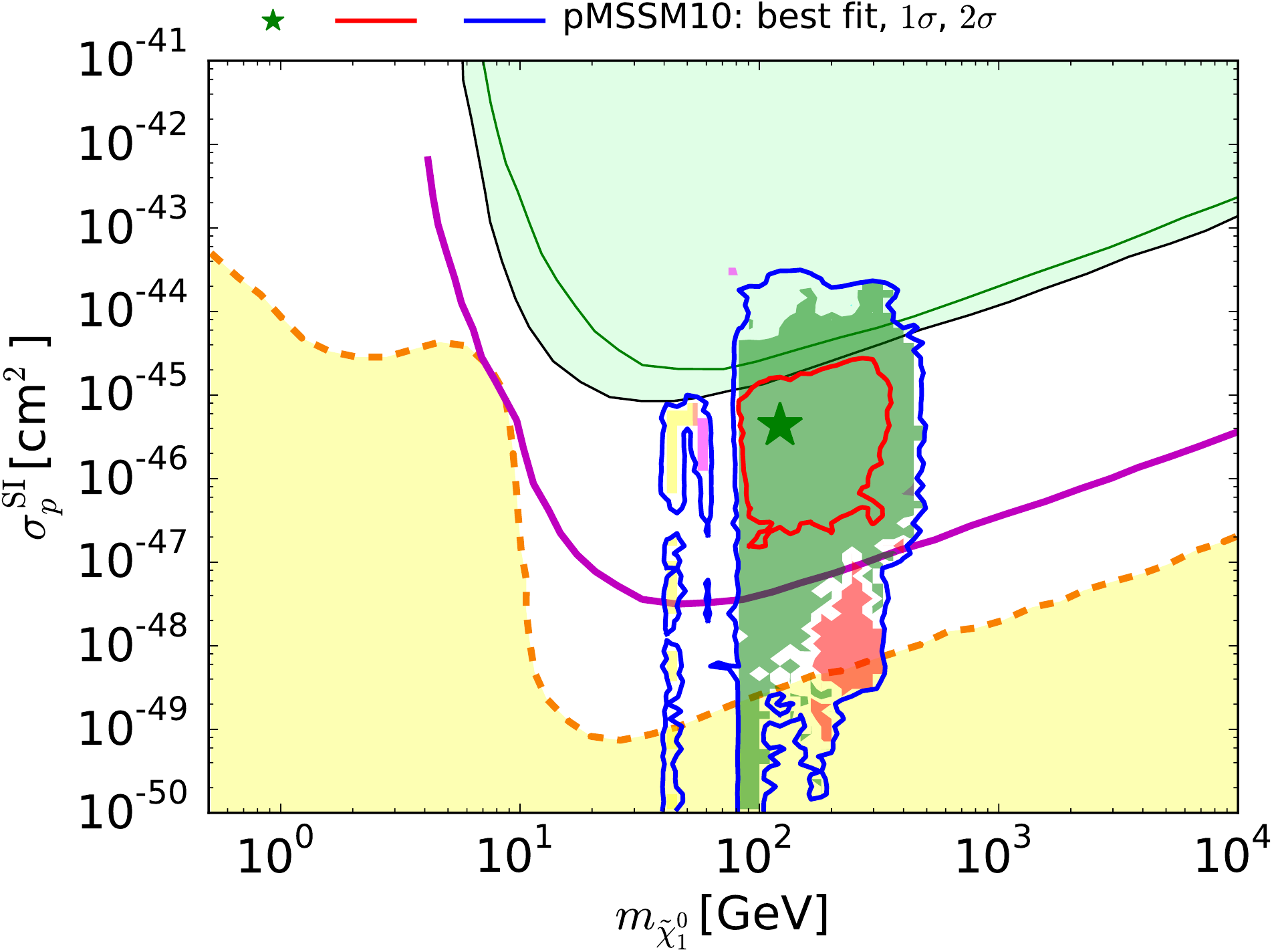}}\\
\resizebox{15cm}{!}{\includegraphics{n12c_dm_legend}}
\end{center}
\vspace{-0.5cm}
\caption{\it The $(\mneu1, \ssi)$ planes in the CMSSM (upper left),
the NUHM1 (upper right), the NUHM2 (lower left) and the pMSSM10 (lower right).
The red and blue solid lines are  the $\Delta \chi^2 = 2.30$ and 5.99 contours,
and the solid purple lines show the projected 95\% exclusion sensitivity of the LUX-Zepelin
(LZ) experiment~\protect\cite{LZ}. The green and black lines show the current sensitivities of the XENON100~\protect\cite{Xe100}
and LUX~\protect\cite{LUX} experiments,
respectively, and the dashed orange line shows the
astrophysical neutrino `floor'~\protect\cite{Snowmass}, {below which astrophysical neutrino backgrounds dominate (yellow region)}.
}
\label{fig:ssi}
\end{figure*}

We now turn to the capabilities of direct DM search experiments to cast
light on the various DM mechanisms. Fig.~\ref{fig:ssi} displays the
$(\mneu1, \ssi)$ planes for the CMSSM (upper left), the NUHM1 (upper right),
the NUHM2 (lower left) and the pMSSM10 (lower right), where \ssi\ is the cross section for
spin-independent scattering on a proton. {Our computation of $\ssi$ follows the procedure
described in~\cite{MC9}, and we have once again adopted for the $\pi$-nucleon $\sigma$ term
the value $\Sigma_{\pi N} = 50 \pm 7$ MeV.}  As previously, the
$\Delta \chi^2 = 2.30$ and 5.99 contours are shown as red and blue lines.
The sensitivities of the XENON100~\cite{Xe100} and LUX~\cite{LUX} experiments are shown as green and
black lines, respectively, and the prospective
sensitivity of the LUX-Zepelin (LZ) 
experiment~\cite{LZ} is shown as a purple line: {the projected sensitivity of the XENON1T 
experiment~\cite{xenon1t} lies between the current LUX bound
and the future LZ sensitivity.} 
Also shown, as a dashed orange line, is the
neutrino `floor', below which astrophysical neutrino backgrounds would dominate
any DM signal~\cite{Snowmass} {(yellow region)}.

In the CMSSM case, we see that the current XENON100 and LUX data already put strong pressure
on models where the focus-point or $\cha{1}$ coannihilation mechanism dominates. 
There are borderline regions that are formally
excluded by the \ssi\ data considered in isolation, but become permitted at the 95\% CL in a global fit including other
observables, and also
{due to the uncertainties in the calculation of \ssi\ that
have been included in the evaluation of the global $\chi^2$ function~\cite{MC9}}. 
We also see that the {$\cha1$ coannihilation region and most of the $H/A$ funnel region} would be accessible to
the planned LZ experiment. However, much of the ${\staue}$ coannihilation region lies below the LZ sensitivity,
though it could be accessible to a 20-tonne DM experiment such as
Darwin~\cite{Darwin}. {However, this region can be covered in the
complementary direct searches at the LHC, as discussed previously.}
In the case of the NUHM1, 
the $\cha{1}$ coannihilation and $H/A$ funnel regions still lie largely within the LZ range,
and the pure $H/A$ funnel region is also within reach of LZ.
However, the ${\staue}$ coannihilation and hybrid regions
extend below the reach of LZ and Darwin, even below the neutrino
`floor'. {Again, this region could be partially covered by the
complementary LHC direct searches.}
Similar qualitative conclusions apply to the
NUHM2, with the additional observation that the small ${\tilde t_1}$ coannihilation regions
lie below the LZ sensitivity
and straddle the neutrino `floor'.

Finally, we see that whereas the region of the pMSSM10 parameter space that is
favoured at the 68\% CL lies within reach of the LZ experiment, 
{as is the case for much of the $\cha1$ coannihilation region},
there are models in the $\cha1$ and ${\staue}$ coannihilation regions,
as well as in the $h$ and $Z$ funnels {and uncoloured regions where none of these mechanisms dominate, in which} cancellations in the
spin-independent matrix element may drive \ssi\ below the neutrino `floor'.
{It should be kept in mind here (see the discussion in \cite{MC11}), that
  these very low values of \ssi\ are due to cancellations~\cite{Cancellations} between
different contributions to the matrix element for spin-independent scattering on protons. In \cite{MC11} it was shown that similar
cancellations hold when the cross section for spin-independent
scattering on neutrons is considered, instead of the proton case shown in
\reffi{fig:ssi}.}

{Table~\ref{tab:detectability} also summarizes the observability of
DM particles in direct searches in the different scenarios considered.
We see a degree of complementarity between the LHC and direct DM searches.
}

{We have focused in this article on the prospects for direct searches for
DM scattering. A complementary probe
of the properties of supersymmetric DM is through indirect
detection, searching for the traces of DM annihilation in the
Galaxy. A number of recent works have focussed on
this. For example, \cite{Cahill-Rowley:2014boa} has demonstrated that Fermi-LAT
satellite limits on $\gamma$-ray emission in dwarf spheroidal
galaxies~\cite{Ackermann:2013yva} do not currently affect the
parameter space of the pMSSM, although they may do so in the
future~\footnote{{Ref.~\cite{Ackermann:2015zua} appeared while
this work was being finalized, and may be used to constrain the $Z$- and $h$-funnel regions of the pMSSM10.
We plan on incorporating this into a
future analysis.}}. Constraints from IceCube and the HESS telescope
have been investigated in~\cite{Fowlie:2013oua,Roszkowski:2014iqa}. 
IceCube limits~\cite{Aartsen:2012kia}  do not currently affect the
pMSSM10 parameter space, while HESS bounds~\cite{Abramowski:2013ax}
are primarily on pure wino states, which must have masses greater than a
TeV~\cite{Cohen:2013ama,Fan:2013faa} in order to provide a thermal relic. Since the mass of the lightest
neutralino is low in our models due to the incorporation of the
$(g-2)_{\mu}$ constraint, this does not affect our fits. Accordingly
we do not include these data sets in this work, but we plan on implementing likelihoods for these searches in future analyses.}


\section{Summary and Conclusions}

We have analyzed in this paper the mechanisms that play dominant roles in bringing the relic
neutralino density into the range allowed by cosmology in the CMSSM, the NUHM1, the NUHM2
and the pMSSM10. We have delineated the regions of the parameter spaces of these models
where dominant roles are played by $\stau1$, $\stopone$ or $\cha1$ coannihilation, or funnel regions where
the neutralino annihilates rapidly via the heavy Higgs bosons $H/A$, and also {regions}
where the neutralino has a significant Higgsino component. In the CMSSM, the NUHM1 and the NUHM2
we find that different mechanisms operate in different regions of the parameter spaces, with
relatively small hybrid regions where two mechanisms contribute. {In the pMSSM10,
$\cha1$ coannihilation dominates in most of the parameter space, with some contributions from
other processes in specific ragions.}

{Our assessments of the observability of supersymmetry within different
models, depending on the dominant DM mechanisms, are summarized in Table~\ref{tab:detectability}.}
Within the CMSSM, the NUHM1 and the NUHM2, $\ETslash$ searches at the LHC can explore
significant portions of the $\stau1$ coannihilation regions. These regions also offer the
possibility that the $\stau1$ may be relatively long-lived, and detectable at the LHC as a
long-lived charged particle. {There} are regions of the NUHM2 parameter space where
$\stopone$ coannihilation dominates, which can also be explored by $\ETslash$ searches
at the LHC. The $\cha1$ coannihilation and focus-point regions of these models can be
explored by the LUX-Zepelin direct DM search experiment.
Much of the $H/A$ funnel regions in these models may be explored
via LHC searches for the heavy Higgs bosons, and also via direct DM searches with the
LUX-Zepelin and Darwin experiments. On the other hand, the $\stau1$ coannihilation regions
seem likely to lie beyond the reaches of the direct DM searches.

Within our analysis of the pMSSM10, $\cha1$ coannihilation is the
dominant DM mechanism in {most of} the parameter space, {though this
might change with a different set of independent pMSSM parameters}. {In
addition, there are regions where ${\staue}$ coannihilation 
or direct-channel annihilation via a $h$ and $Z$ funnel may dominate.}
Parts of the pMSSM10 model space can be explored at the LHC
via $\ETslash$, $H/A$ and other searches, and parts by direct DM searches.
However, we find no long-lived particle signature in the region of the pMSSM10
parameter space that is currently favoured statistically. 
Overall, large parts of the pMSSM10
parameter space could escape the LHC searches considered here,
{but a large fraction would be accessible to future DM experiments}.

Our analysis shows that the LHC and direct matter searches offer significant
prospects for discovering SUSY if it is responsible for the cosmological
CDM, and {in many cases} the mode of discovery can reveal the nature of the
dominant mechanism responsible for determining the CDM density.
We look forward with interest to learning what the LHC and direct searches
will be able to tell us about SUSY DM.


\subsubsection*{Acknowledgements}

The work of  O.B., J.E., S.M., K.A.O., K.S. and K.J.de~V. is supported in part by
the London Centre for Terauniverse Studies (LCTS), using funding from
the European Research Council 
via the Advanced Investigator Grant 267352.
The work of R.C. is supported in part by the National Science Foundation under 
Grant No. PHY-1151640 at the University of Illinois Chicago and in part by Fermilab, 
operated by Fermi Research Alliance, LLC under Contract No. De-AC02-07CH11359 
with the United States Department of Energy.
This work of M.J.D.
is supported in part by the Australia Research Council. The work of J.E. is also supported in part by STFC
(UK) via the research grant ST/L000326/1, and
the work of H.F. is also supported in part by STFC (UK).
The work of S.H. is supported 
in part by CICYT (grant FPA 2013-40715-P) and by the
Spanish MICINN's Consolider-Ingenio 2010 Program under grant MultiDark
CSD2009-00064. The work of D.M.-S. is supported by the European Research Council 
via Grant BSMFLEET 639068.
The work of K.A.O. is supported in part by DOE grant
DE-SC0011842 at the University of Minnesota. The work of G.W.\ is supported in 
part by the Collaborative Research Center SFB676 of the DFG, ``Particles, Strings and the early Universe", 
and by the European Commission through the ``HiggsTools" Initial Training Network PITN-GA-2012-316704. 


\end{document}